\documentclass[twocolumn,showpacs,preprintnumbers,amsmath,amssymb]{revtex4}

\usepackage{graphicx}
\usepackage{dcolumn}
\usepackage{bm}

\begin{document}


\title{Measurement of the Absolute Differential Cross Section for np Elastic
Scattering at 194 MeV}

\author{M. Sarsour$^{1}$} \altaffiliation[Present address:]{Dept. of Physics, Texas
A \& M University, College Station, TX, USA}
\author{T. Peterson$^{1}$} \altaffiliation[Present address:
]{Dept. of Radiology and Radiological Sciences, Vanderbilt University, Nashville, TN, USA}
\author{M. Planinic$^{1}$}
\altaffiliation[Present address: ]{Dept. of Physics, University of Zagreb, Zagreb,
Croatia}
\author{S.E. Vigdor$^{1}$}
\author{C. Allgower,$^{1}$}
\author{B. Bergenwall$^{2}$}
\author{J. Blomgren$^{2}$}
\author{T. Hossbach$^{1}$}
\author{W.W. Jacobs$^{1}$}
\author{C. Johansson$^{2}$}
\author{J. Klug$^{2}$}
\author{A.V. Klyachko$^{1}$}
\author{P. Nadel-Turonski$^{2}$} \altaffiliation[Present address: ]{Dept. of Physics, George
Washington University, Washington, DC, USA},
\author{L. Nilsson$^{2}$}
\author{N. Olsson$^{2}$}
\author{S. Pomp$^{2}$}
\author{J. Rapaport$^{3}$}
\author{T. Rinckel$^{1}$}
\author{E.J. Stephenson$^{1}$}
\author{U. Tippawan$^{2,4}$}
\author{S.W. Wissink$^{1}$}
\author{Y. Zhou$^{1}$\\}

\affiliation{$^{1}$Indiana University Cyclotron Facility,
Indiana University, Bloomington, IN, USA\\
$^{2}$Uppsala University, Uppsala, Sweden\\
$^{3}$Ohio University, Athens, OH, USA\\
$^{4}$Chiang Mai University, Chiang Mai, Thailand\\
}

\date{\today}

\begin{abstract}
A tagged medium-energy neutron beam has been used in a precise measurement of the absolute
differential cross section for np back-scattering.  The results resolve significant
discrepancies within the np database concerning the angular dependence in this regime. The
experiment has determined the absolute normalization with $\pm 1.5\%$ uncertainty,
suitable to verify constraints of supposedly comparable precision that arise from the rest
of the database in partial wave analyses.  The analysis procedures, especially those
associated with the evaluation of systematic errors in the experiment, are described in
detail so that systematic uncertainties may be included in a reasonable way in subsequent
partial wave analysis fits incorporating the present results.
\end{abstract}

\pacs{13.75.Cs, 21.30.-x, 25.10.+s}
\keywords{neutron-proton elastic scattering; precise absolute differential
cross sections; tagged neutron beam}
\maketitle

\section{INTRODUCTION}

\bigskip

Theoretical treatments and applications of the nucleon-nucleon (NN) force at low and
intermediate energies have progressed considerably in sophistication through the past
decade.  Partial wave analyses and potential model fits to the NN scattering database have
incorporated explicit allowance for breaking of isospin ($I$) symmetry, \emph{e.g.}, by
removing constraints that previously required equal $I=1$ phase shifts for the pp and np
systems, and have been used to constrain the pion-nucleon-nucleon coupling constant
\cite{review}. Effective field theory approaches \cite{Mach2003} have become competitive
with more traditional meson-exchange models of the interaction, in terms of the quality of
fit provided to the database and the number of adjustable parameters employed, while
holding out the promise of providing internally consistent two- and three-nucleon forces
from the same theory. Striking success has been achieved in \emph{ab initio} calculations
of the structure of light nuclei \cite{abinit} by combining phenomenological three-nucleon
forces with NN interactions taken without modification from fits to the NN scattering
database. An important aspect in these advances has been the approach toward consensus on
which measurements should be included in an NN database to which conventional $\chi^2$
optimization techniques can be sensibly applied.  The rejection of specific, allegedly
flawed, experiments from the database has not been without controversy. In the present
paper, we report detailed results from a new np scattering experiment addressing one of
the most prominent of these controversies.

\begin{figure*}[htp!]
\includegraphics[scale=0.5]{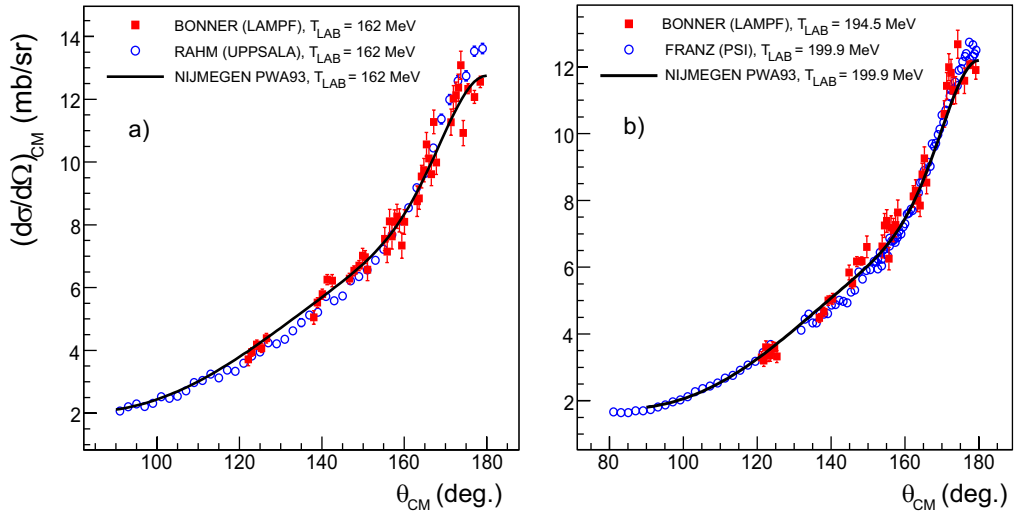}
\caption{\label{fig:xscomp}\small Comparison of previous np scattering differential cross
section measurements (a) from Uppsala \cite{Rah1998} and Los Alamos \cite{Bon1978}, and
(b) from PSI \cite{Fra2000} and Los Alamos \cite{Bon1978} near 200 MeV.  The Los Alamos
data in each case are represented by closed squares and the other data by open circles.
The experimental results are compared to the Nijmegen PWA93 \cite{Sto1993} partial wave
analysis solution evaluated at appropriate energies. The Los Alamos data have been
renormalized by factors of 1.092 in (a) and 1.078 in (b) to bring them into agreement with
the PWA. The relative cross sections reported in \cite{Fra2000} have been similarly
normalized here, while the reported absolute cross section scale for the Uppsala data has
been retained.}
\end{figure*}

Discrepancies among different experiments have led to a drastic pruning of cross section
measurements for intermediate-energy np scattering.  For example, the SAID partial wave
analysis (PWA) of the np database \cite{SAID} rejects more than 40\% of all measured cross
sections in the range 100-300 MeV in neutron laboratory kinetic energy.  The rejected
fraction is even larger in the Nijmegen PWA \cite{Sto1993,Ren2001}, especially so for
scattering at center-of-mass angles beyond 90$^\circ$.  The rejected data include nearly
all of the most recent experiments, carried out by groups at Uppsala \cite{Rah1998} and
Freiburg \cite{Hur1980,Fra2000}.  The problems are illustrated in Fig.~\ref{fig:xscomp} by
the comparison of data from these two groups with earlier Los Alamos measurements
\cite{Bon1978} that dominate the medium-energy back-angle cross section data retained in
the database.  Clear differences among these data sets are seen in the shape of the
angular distribution.  Other differences, reflecting the general experimental difficulty
in determining the absolute scale for neutron-induced cross sections, are masked in the
figure by renormalization factors that have been applied in the partial wave analyses.
Removal of the Uppsala and Freiburg data, which exhibit fairly similar angular
dependences, begs the question of whether the $\chi^2$ criterion used to reject them
\cite{Sto1993,Bug1995,Ren2001,SAID} may subtly bias the PWA results toward agreement with
older measurements that might have had their own unrecognized systematic errors.

The np back-angle cross section discrepancies have been highlighted in debates concerning
the value and extraction methods for the charged $\pi$NN coupling constant $f^2_c$ (in the
notation of pseudovector formulations of the interaction, or equivalently
$g_{\pi^\pm}^2/4\pi$ in pseudoscalar formulations) \cite{Eri1995,PiN2000}. np scattering
PWA's appear to determine this basic parameter of the NN force well: \emph{e.g.}, the
Nijmegen analysis \cite{Sto1993} yields $f_c^2 = 0.0748 \pm 0.0003$ (equivalent to
$g_{\pi^\pm}^2/4\pi = 13.54 \pm 0.05$), and the authors claim that the constraints are
imposed by the entire database, with no particularly enhanced sensitivity to any specific
observable \cite{Sto1993}. In contrast, Ericson \emph{et al.} have extracted a
significantly higher coupling constant, consistent with older ``textbook" values
($g_{\pi}^2/4\pi \approx 14.4$), by applying controversial pole extrapolation techniques
to the Uppsala back-angle np scattering cross sections alone \cite{Eri1995}. While much
debate has centered on the rigor of the pole extrapolation method
\cite{PiN2000,Ren1998,Eri1998}, it is clear that the discrepancy in coupling constant
values arises in large part \cite{Arn1995} from the cross section discrepancies between
the Uppsala measurements and the ``accepted" database. An experimental resolution of these
discrepancies is highly desirable, especially if a new experiment can also pin down the
absolute cross section scale. Bugg and Machleidt have pointed out \cite{Bug1995} that the
largest uncertainty in their determination of $f^2_c$ is associated with the normalization
of np differential cross sections, which are often allowed to float from the claimed
normalization in individual experiments by 10\% or more in PWA's. In contrast, the
Nijmegen group claims \cite{Ren2001} that, despite sizable normalization uncertainties in
existing elastic scattering data, precise total cross section measurements fix the np
absolute cross section scale to $\pm 0.5\%$ accuracy.  This claim could also be checked by
a new experiment that provides good experimental precision on absolute differential cross
sections.

In the present paper we report detailed results from such a new experiment, designed to
resolve these np back-angle cross section discrepancies.  The experiment employed
techniques completely independent of those used in other medium-energy measurements, in
order to provide tight control over systematic errors. A tagged neutron beam
\cite{Pet2004} centered around 194 MeV kinetic energy bombarded carefully matched,
large-volume CH$_2$ and C targets, which permitted accurate subtraction of backgrounds
from quasifree scattering and other sources.  The bombarding energy range was chosen to
match approximately that used in earlier high-precision np scattering polarization
measurements from the Indiana University Cyclotron Facility (IUCF)
\cite{Vig1992,Sow1987,Bow1994}. Recoil protons from np scattering were identified in a
detector array of sufficient angular coverage to measure the differential cross section at
all c.m. angles beyond 90$^\circ$ simultaneously.

The tagging allows accurate determination of the absolute scattering probability for the
analyzed subset of all neutrons incident on a secondary target, but it also offers a host
of other, less obvious, advantages important to a precise experiment: (1) accurate
relative normalization of data taken with CH$_2$ \emph{vs.} C targets; (2) event-by-event
determination of neutron energy, impact point and incidence angle on the secondary target,
with the latter measurement being especially important for cross section measurements very
near 180$^\circ$ c.m. scattering angle; (3) three-dimensional location of background
sources displaced from the secondary target \cite{Pet2004}; (4) precise measurement of the
detector acceptance for np scattering events; (5) methods to tag np scattering event
subsamples that should yield identical cross section results but different sensitivity to
various sources of systematic error. The tagging was thus essential to the entire approach
of the experiment; no \emph{extra} work was required to extract \emph{absolute} cross
sections, and thereby to provide an important calibration standard for medium-energy
neutron-induced reactions.

The basic results of this experiment have recently been reported briefly \cite{Sar2005}.
In the present paper we provide more detail on the comparison of results to PWA's, on the
data analysis procedures and on the evaluation and characterization of systematic
uncertainties.  Such details are important for resolving the sort of discrepancies that
have plagued the np database. Partial wave analyses should, in principle, incorporate
experimental systematic as well as statistical errors in optimizing fits to data from a
wide variety of experiments.  In order to do so, they must have access to clear
delineations of which errors affect only the overall normalization, which have
angle-dependence and, in the latter case, what the degree of correlation is among errors
at different angles.  Overall systematic uncertainties in the absolute cross sections
reported here average $\pm 1.6\%$, with a slight angle-dependence detailed herein.
Statistical uncertainties in the measurements are in the range $\pm (1-3)\%$ in each of 15
angle bins.

\section{DETAILS OF THE EXPERIMENT}

\begin{figure}[htp!]
\includegraphics[scale=0.4]{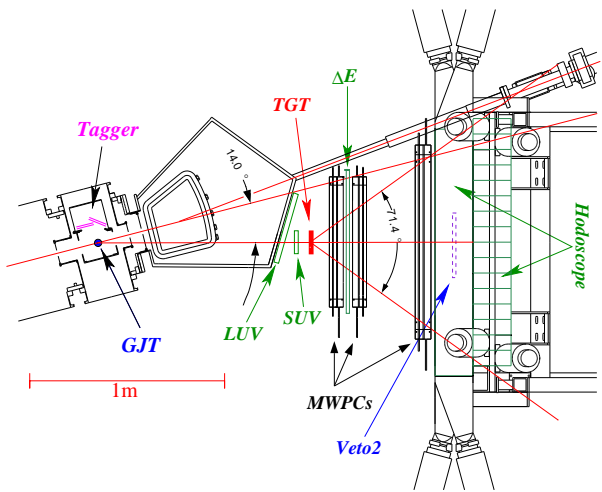}
\caption{\label{fig:setup}\small Top view of the np scattering experiment setup.}
\end{figure}

The experiment was carried out in the IUCF Cooler ring \cite{Pol1991}, with apparatus (see
Fig.~\ref{fig:setup}) installed in a ring section where the primary stored proton beam was
bent by 6$^\circ$. A primary electron-cooled unpolarized proton beam of 202.5 MeV kinetic
energy and typical circulating current of 1--2 mA was stored in the ring. Neutrons of
185-197 MeV were produced via the charge-exchange reaction p+d $\rightarrow$ n+2p when the
proton beam passed through a windowless internal deuterium gas jet target (GJT) of typical
thickness $\approx 3 \times 10^{15}$ atoms/cm$^2$. The ultra-thin target permitted
detection of the two associated low-energy recoil protons from the production reaction in
double-sided silicon strip detectors (DSSD's) comprising the ``tagger".  Measurements of
energy, arrival time and two-dimensional position for both recoil protons in the tagger,
when combined with the precise knowledge of cooled primary proton beam direction and
energy, allowed four-momentum determination for each tagged neutron on an event-by-event
basis.  During the measurement periods, the stored proton beam was operated in ``coasting"
mode, with rf bunching turned off to minimize the ratio of accidental to real two-proton
coincidences in the tagger.  The proton beam energy was then maintained by velocity
matching (induced naturally by mutual electromagnetic interactions) to the collinear
electron beam in the beam cooling section of the ring.

Details of the layout, design and performance of the tagger detectors and of the forward
detector array used to view np scattering events from the secondary target are provided in
Ref. \cite{Pet2004}.  Here, we summarize the salient features briefly.  The tagger
included an array of four 6.4$\times$6.4 cm$^2$ DSSD's with 480 $\mu$m readout pitch in
two orthogonal ($x^\prime$,$y^\prime$) directions, each followed by a silicon pad
(``backing") detector (BD) of the same area. The DSSD's were positioned about 10 cm away
from the center of the gas jet production target.  Each DSSD had 128 $x^\prime$ and 128
$y^\prime$ readout channels. Readout was accomplished with front-end application-specific
integrated circuits (ASIC's) that provided both timing and energy information
\cite{Pet2004}. The timing signals provided to external electronics consisted of the
logical OR over groups of 32 adjacent channels of leading-edge discriminator signals based
on fast shaped and amplified analog signals generated in the ASIC's.  The timing signals
available from $4~x^\prime~\times~4~y^\prime$ logical pixels for each DSSD permitted
operation of the tagger in a self-triggering mode, where the time-consuming digitization
of slow pulse height signals from all 1024 DSSD channels could be initiated by logic based
solely on the tagger hit pattern, as reconstructed from the fast timing signals. This
self-triggering was critical to the determination of precise absolute cross sections,
because it allowed acquisition of data to count directly the flux of tagged neutrons that
did not interact in the secondary target or in any of the forward detectors.

Only recoil protons that stopped either in the DSSD's ($E_p \lesssim 7$ MeV) or BD's ($E_p
\lesssim 11$ MeV) were considered in the data analysis, because for these the tagger
provided a measurement of total kinetic energy with good resolution.  By combining these
energy measurements with position measurements for both recoil protons, we were able to
determine the energy and angle of each tagged neutron, within their broad distributions,
with respective resolutions of $\sigma_E \approx 60$ keV and $\sigma_{angle} \approx$ 2
mrad.  As part of this determination, we reconstructed the longitudinal origin
($z_{vertex}$) of each produced neutron within the extended GJT density profile with a
resolution of $\approx$ 2 mm, by comparing neutron momentum magnitudes inferred by
applying energy conservation (independent of vertex position) \emph{vs.} vector momentum
conservation (dependent on vertex position) to the tagger information for the two recoil
protons. Similar resolution was obtained for the transverse coordinates at which each
tagged neutron impinged upon a secondary scattering target (TGT in Fig.~\ref{fig:setup})
positioned 1.1 m downstream of the GJT.

Two solid secondary targets were used during the production running: a $20 \times 20
\times 2.5$ cm$^3$ slab of polyethylene (CH$_2$) containing $1.99 \times 10^{23}$ hydrogen
atoms/cm$^2$ and a graphite target of known density machined to have identical transverse
dimensions and the same number of carbon atoms per unit area. Each target thickness was
determined to $\pm 0.4\%$ by weighing. Data were collected in 18-hour cycles, comprising 6
hours of running with the CH$_2$ target, followed by 6 hours with C and 6 more hours with
CH$_2$. The frequent interchange of the targets facilitated accurate background
subtractions. Both targets intercepted neutrons over an approximate production angle range
of $14^\circ \pm 5^\circ$, and cuts were generally placed on the tagger information during
data analysis to confine attention to tagged neutrons that would hit the secondary target.
Such tagged neutrons were produced at a typical rate of $\sim$ 200 s$^{-1}$, leading to
typical free np back-scattering (angle-integrated) rates $\sim$ 1 s$^{-1}$ from the CH$_2$
target.

Protons emerging from the secondary target were detected in a forward array of plastic
scintillators for triggering and energy information, and a set of (three-plane) multi-wire
proportional chambers (MWPCs) for tracking, as indicated in Fig.~\ref{fig:setup}.  The
plastic scintillators included large upstream veto (LUV) and small upstream veto (SUV)
counters to reject charged particles produced upstream of the secondary target. The
$\Delta$E scintillator was separated from the secondary target by a MWPC to permit easy
discrimination against np scattering events initiated in that scintillator. The rear
hodoscope comprised 20 plastic scintillator bars \cite{Geneva} of sufficient thickness (20
cm) to stop 200 MeV protons and give 15-20\% detection efficiency for 100-200 MeV
neutrons.  All forward detectors were rectangular in transverse profile, with the rear
MWPC and hodoscope spanning a considerably larger vertical than horizontal acceptance. The
entire forward array provided essentially 100\% ($> 50\%$) geometric acceptance for np
scattering events initiated at the CH$_2$ target for angles $\theta_{c.m.} \gtrsim
130^\circ$ ($\theta_{c.m.} \gtrsim 90^\circ$).  For c.m. angles forward of 90$^\circ$ the
large size and significant neutron detection efficiency of the hodoscope provided a small
efficiency for detecting forward-scattered neutrons in coincidence with larger-angle
protons that fired at least the first two MWPCs.

The tagger and forward detector array were designed to facilitate a kinematically complete
double-scattering experiment with a first target giving negligible energy loss.  With the
same apparatus, a similar measurement of pp scattering was possible simultaneously.  For
this purpose one could use the tagger to detect a single large-angle recoil deuteron
instead of two recoil protons, in order to tag a secondary proton beam via pd elastic
scattering in the GJT.  By requiring a coincidence between a single hit in the tagger and
a signal from the small upstream veto scintillator (SUV in Fig.~\ref{fig:setup}), we could
define a secondary proton beam of very similar transverse dimensions to the tagged neutron
beam. Another scintillator (Veto2) placed just in front of the rear hodoscope allowed us
to distinguish, at trigger level, between protons from pd elastic scattering that
traversed the forward array without further nuclear interactions and protons that
scattered out of this secondary beam in material following SUV.  In the present paper, we
discuss only the former group, as their yield provides an accurate relative normalization
of runs taken with the CH$_2$ \emph{vs.} C targets.

The triggered events of interest for the present analysis were recorded in four mutually
exclusive event streams, three for tagged neutron candidates (consistent with two distinct
tagger hits and no accompanying signals from LUV or SUV) and one for tagged proton
candidates (consistent with a single tagger hit in prompt coincidence with both LUV and
SUV).  The trigger logic defined these event streams as follows: (1) tagged neutrons with
no rear hodoscope coincidence, providing a prescaled (by a factor of 20) sample for
neutron flux monitoring; (2) $np$ scattering candidates for which a tagged neutron was in
coincidence with signals from both the $\Delta$E scintillator and the rear hodoscope; (3)
tagged neutrons in coincidence with the hodoscope but \emph{not} with the $\Delta$E
scintillator, a sample used for evaluating the neutron detection efficiency of the
hodoscope \cite{Pet2004}; (4) tagged protons in coincidence with both the $\Delta$E and
Veto2 scintillators, providing a prescaled (by a factor of 10) sample including pd elastic
scattering events from the GJT, used to cross-normalize C and CH$_2$ secondary target
runs.  The total flux of tagged neutrons intercepting the secondary target was determined
from a sum over event streams (1)+(2)+(3), while comparative analyses of the three streams
facilitated crosschecks to calibrate the system \cite{Pet2004} and aid understanding of
potential systematic errors.

\section{DATA ANALYSIS}

\subsection{Cuts and conditions on tagged neutron beam properties}

The general philosophy of the data analysis was to define properties of the tagged neutron
beam by identical cuts applied to event streams (1)-(3), so that associated systematic
uncertainties would cancel in the yield ratios from which the absolute np scattering cross
section is extracted.  Among these common cuts, described in more detail below, were ones
to remove BD noise contributions correlated among the four quadrants of the tagger, to
identify the recoil particles detected in the tagger, and to divide the tagged neutron
events into subsamples for subsequent analysis. Additional cuts defined a fiducial range
for the tagged neutron's predicted transverse coordinates at the secondary target ($\mid
x_{tag} \mid <$9.5 cm and $\mid y_{tag} \mid <$9.5 cm), and selected prompt tagger
two-particle coincidences ($|t_{p1} - t_{p2} - 30~{\rm ns}| \leq 70$ ns, where
$t_{p1}~(t_{p2})$ is the arrival time of the recoil proton with the larger (smaller) DSSD
energy deposition). Software cuts applied to event stream (2) alone to identify free np
scattering events were kept to a minimum in order to avoid complicated systematic errors.
We relied instead on the accuracy of the background subtractions, which could be verified
to high precision. Before application of cuts, additional MWPC requirements were added in
software to amplify the hardware definitions of the various triggers.  Thus, at least one
hit in the $x$-plane and at least one hit in the $y$-plane were required for each of the
three MWPC's for events from stream 2.

\subsubsection{Particle identification}

The correlation of DSSD vs. BD energy depositions was used to select two basic event
classes for analysis of each of the three tagged neutron event streams: (a) ``2-stop''
events, where both protons associated with the neutron stopped inside the DSSD (either the
same or different quadrants of the tagger); and (b) ``1-punch'' events, where one of the
protons stopped inside a DSSD and the other punched through to the BD in a different
quadrant and stopped there.  These two classes, as discussed further below, differ
significantly in neutron energy ($E_n$) and position ($x_{tag}$) profiles, allowing an
important crosscheck on the accuracy of the tagging technique by comparing np cross
sections extracted independently from each class.  The 2-stop events were further
subdivided according to whether the higher of the two recoil proton DSSD energy
depositions ($E_{p1}$) was below or above 5.0 MeV.  Protons with $E_{p1} > 5.0$ MeV range
out near the exit of the DSSD, and hence possibly in dead layers at the back of the DSSD
or front of the BD, making this event class subject to somewhat greater ambiguity
regarding complete recoil proton energy reconstruction and accuracy of the predicted
$x_{tag}$ value for the tagged neutron.  Events where both protons punched through to the
BD's, or where either punched through the BD itself, were not included in the analysis
because they corresponded to $E_n$ below the range of interest.

\begin{figure}[htb!]
\begin{center}
\includegraphics[scale=0.2]{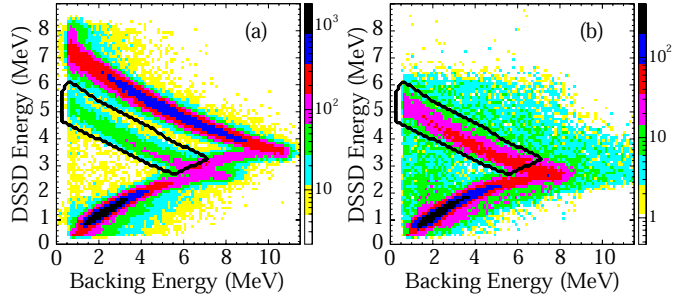}
\caption{\label{fig:pid}\small Particle identification plot for one DSSD-BD combination
for (a) tagged protons and (b) tagged neutrons.}
\end{center}
\end{figure}

Figure~\ref{fig:pid} shows raw spectra for both tagged neutrons (event stream 1 in frame
(b)) and tagged protons (event stream 4 in frame (a)) of the energy deposited in a typical
DSSD quadrant vs. that in the companion BD when the latter is non-zero. The tagged proton
events exhibit clear recoil proton (lower) and deuteron (upper) particle identification
loci, while only the proton locus remains for tagged neutron events.  The loci bend
backwards when the detected particle begins to punch through the BD.  The two-dimensional
gate (dark boundary) shown in each frame was used to select recoil protons that enter and
stop inside the BD, \emph{e.g.}, to identify the 1-punch tagged neutron events.  Note that
the most intense region along the proton locus, corresponding to deuteron breakup events
with an energetic large-angle proton, is thereby eliminated.  So are events lying off the
proton locus, where the backing detector response may be corrupted by noise or pileup.

\begin{figure*}[htp!]
\begin{center}
\includegraphics[scale=0.3]{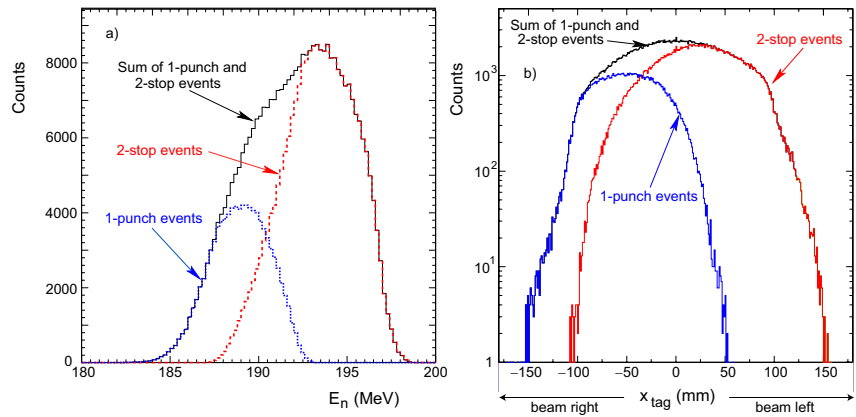}
\caption{\label{fig:enxtag}\small The reconstructed energy (a) and horizontal transverse
coordinate (b) of tagged neutrons at the secondary target for 1-punch events, 2-stop
events and their sum (solid black line).}
\end{center}
\end{figure*}

Figure~\ref{fig:enxtag} shows the reconstructed $E_n$ and $x_{tag}$ distributions for the
tagged neutrons in the 2-stop (for all values of $E_{p1}$) vs. 1-punch samples.  While the
two samples yield overlapping distributions, it is clear that the 1-punch events
correspond on average to lower-energy neutrons at larger production angles (preferentially
populating the beam-right side of the secondary target).

\subsubsection{Correlated noise in the BD}

\begin{figure}[htb!]
\begin{center}
\includegraphics[scale=0.42]{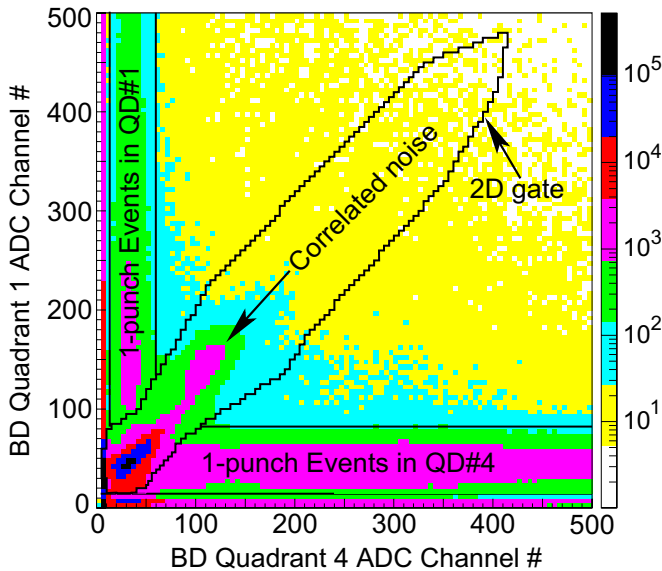}
\caption{\label{fig:cornoise} \small The ADC outputs of two BD's plotted against one
another, revealing regions of 1-punch tagged neutron events, but also a region dominated
by correlated noise in both quadrants.  Events along the $x$- and $y$-axes, where one BD
signal falls in the pedestal region, are also mostly valid 1-punch events.}
\end{center}
\end{figure}

Special care was taken in the definitions of 1-punch and 2-stop events to minimize effects
of substantial detector noise picked up by the large-capacitance BD's. An important source
of this noise was discovered to arise from the initiation of pulse height information
readout on the adjacent DSSD front-end electronics \cite{Pet2004}. The induced noise was
strongly correlated among the four BD's, as illustrated in Fig.~\ref{fig:cornoise}. This
figure reveals two uncorrelated bands parallel to the $x$ and $y$ axes, due to 1-punch
events in one of the quadrants and low pulse-height noise in the other (the pedestals for
each BD appear in ADC channel $\approx$10). But one also observes a strong diagonal band
indicative of noise correlations between the two quadrants.

Since the noise correlation pattern extends beyond a reasonable software threshold, it was
necessary to use a two-dimensional gate, such as that shown in Fig.~\ref{fig:cornoise}, to
bound the noise correlation region.  Candidates for valid 1-punch events were then
required to: (1) surpass a threshold ADC channel ($\approx$15) on at least one BD; (2)
fall outside the noise correlation gates for all BD pairs; (3) not surpass the BD noise
peak (ADC $\approx$70) in more than one BD; and (4) fall within the PID gate in
Fig.~\ref{fig:pid} for the appropriate quadrant. These conditions and the complementary
ones required for 2-stop events reduced the flux of tagged neutrons considered for
subsequent analysis, by removing events with BD pulse height ambiguities, but since they
were applied equally to all tagged neutron event streams, they did not introduce
systematic errors in the np cross section extraction.

\subsection{Corrupted events subtraction}
\label{corrupt}

An event misidentification mechanism discovered during the data analysis was attributed to
an electronics malfunction in the gating or clearing circuit for the electronics module
that was used to digitize the pulse height information for all four BD's. The effect of
the malfunction was to zero out valid BD energy signals for a randomly selected fraction
of punch-through events. The effect was seen clearly, for example, in the pd elastic
scattering events in stream 4, where a software gate placed on the two-body kinematic
correlation between recoil deuteron DSSD energy deposition and forward proton angle could
be used to select events in which the deuteron must have stopped in the BD.  Roughly 3/4
of these events showed the anticipated BD pulse height, but 1/4 had $E_{BD}=0$.  In the
case of tagged neutrons, the corrupted events were misidentified as 2-stop events and gave
systematically incorrect predictions of the tagged neutron trajectory, since some recoil
proton energy was lost.  However, the availability of full information for the surviving
punch-through samples allowed us to emulate the effect and subtract the corrupted events
accurately.

\begin{figure}[htb!]
\begin{center}
\includegraphics[scale=0.42]{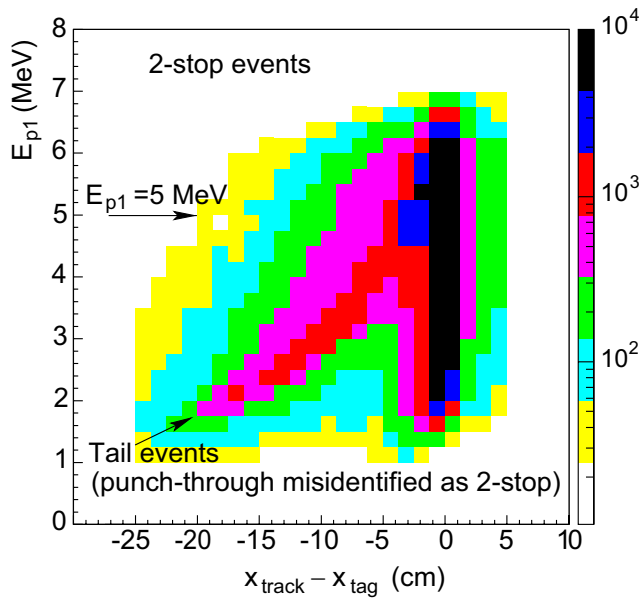}
\caption{\label{fig:tail1}\small The correlation for apparent 2-stop np scattering
candidates between the higher of the DSSD energy depositions for the two recoil protons in
the tagger and the difference in predicted $x$ coordinates at the secondary target from
neutron tagging \emph{vs.} forward proton ray-tracing.  The long correlated tail contains
corrupted punch-through events for which electronic loss of backing detector energy
information has led to misidentification of the event and large systematic errors in the
tagging.}
\end{center}
\end{figure}

The corrupted events were easily distinguished in event stream 2 by using the MWPC
tracking information. Figure~\ref{fig:tail1} shows the correlation for event stream 2
between $E_{p1}$ in the tagger and $x_{track} - x_{tag}$, where $x_{track}$ denotes the
transverse coordinate of the detected proton from np scattering at the secondary target,
as reconstructed from the MWPC hits.  The majority of events have $x_{track} - x_{tag}
\approx 0$, independent of $E_{p1}$, as expected when both the tagging and tracking are
accurate. The corrupted events populate the ``tail'' to the left of the most intense band,
with $x_{tag}$ exceeding $x_{track}$ by an amount that is strongly correlated with the
recoil proton energy deposition in the DSSD: the lower $E_{p1}$, the larger is the lost
$E_{BD}$ and the consequent error in $x_{tag}$.  While the corrupted events could be
eliminated from event stream 2 by a software gate within Fig.~\ref{fig:tail1}, they could
not have been similarly eliminated from event streams 1 and 3, where there was no forward
proton to track.  Hence, it was essential to find a way to subtract these corrupted events
reliably and consistently from all three tagged neutron event streams.

The corrupted events were simulated using all recorded punch-through events that survived
with their BD energy information intact, by reanalyzing these events after artificially
setting $E_{BD}=0$ in the software before tagging reconstruction. The distribution shapes
of the tail events in Fig.~\ref{fig:tail1} with respect to all variables were accurately
reproduced when this simulation was based on \emph{all} events in Fig.~\ref{fig:pid}(b),
both inside and outside the two-dimensional gate drawn, and also including events where
\emph{both} recoil protons punched through their DSSD's. In order to determine the
fraction of these punch-through events that was affected by the electronics malfunction,
we relied on a comparison of the subsamples of our simulated events and of the apparent
2-stop events that had valid BD timing information despite having $E_{BD}=0$. Because the
BD noise problems necessitated high thresholds to generate timing signals, these
subsamples populate mostly the far tail in Fig.~\ref{fig:tail1}, corresponding to $E_{p1}
\lesssim 3.5$ MeV (thus, to relatively large BD analog signals).  The corrupted fraction
of punch-through events was in this way determined independently for each of the three
tagged neutron event streams, and found to be identical for the three, within the
statistical precision (typically $\approx 1\%$) available in matching simulated and
recorded corrupted event subsamples.  The fraction varied slightly with time during the
production run, but averaged 23$\%$. The success of this simulation and the persistence of
the corrupted fraction across (both tagged neutron and tagged proton) event streams
provide strong support for our assumption that the malfunction affected a random sample of
punch-through events.  Because the fractional loss of 1-punch events to this corruption
was independent of event stream, there was no residual systematic effect on the extracted
1-punch cross sections, but rather only a slight loss of statistical precision.

\begin{figure}[htb!]
\begin{center}
\includegraphics[scale=0.45]{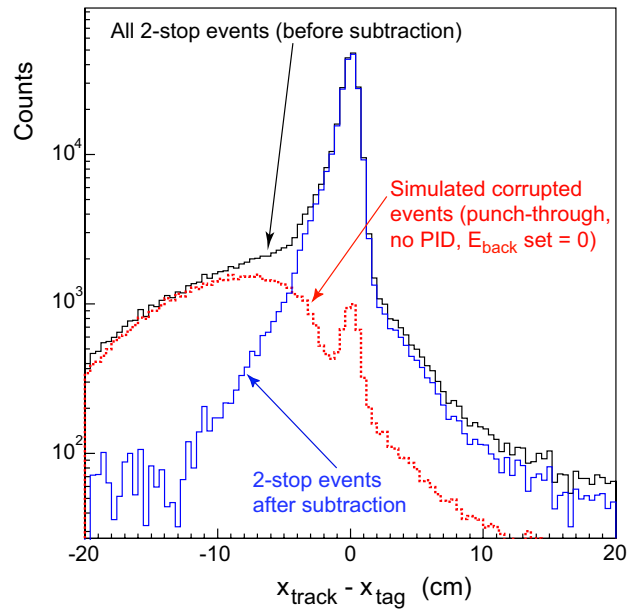}
\caption{\label{fig:tail2}\small The subtraction of the simulated corrupted events removes
the tail from the 2-stop event sample.  The simulated spectrum has been normalized by
matching to the 2-stop subsample characterized by backing detector information with valid
times but zero energy.}
\end{center}
\end{figure}

Figure~\ref{fig:tail2} shows a comparison of the $x_{track} - x_{tag}$ spectrum for all
2-stop events in stream 2 with the simulated corrupted sample, normalized as described
above via the subsample with valid BD timing signals.  The subtraction eliminates
essentially completely the corrupted events with $E_{p1} \lesssim 5.0$ MeV, or $x_{track}
- x_{tag} \lesssim -5$ cm, leaving a reasonably symmetric small background (discussed
further in Sec.~\ref{seqreac}) at $|x_{track} - x_{tag}| > 5$ cm. We therefore assume that
the subtraction is similarly successful for event streams 1 and 3, where we have no
tracking information to compare, and associate a systematic error for the subtraction (see
Sec.~\ref{corsuberror}) that reflects only the uncertainty in the normalization scheme for
the simulated corrupted events.

For $E_{p1}>$ 5.0 MeV, there is a remaining tail of small extent in the subtracted
$x_{track} - x_{tag}$ spectrum in Fig.~\ref{fig:tail2} that arises not from the
electronics malfunction, but rather from recoil protons that barely punch through the
DSSD, while depositing insufficient energy in the BD to be distinguished from noise.
Because of these events, we have separately analyzed the 2-stop samples with $E_{p1} \leq$
5.0 MeV and $E_{p1} >$5.0 MeV.  For the latter sample, after subtracting simulated
corrupted events, we used a two-dimensional software gate on Fig.~\ref{fig:tail1} to
eliminate events in stream 2 that had potentially distorted $x_{tag}$ information, thereby
rejecting 18\% of 2-stop $E_{p1}
>$5.0 MeV events (as opposed to the 23\% of \emph{all} 2-stop events in stream 2 that were
affected by the corruption).  The yields of 2-stop $E_{p1}
>$5.0 MeV events in streams 1 and 3 were then scaled down by the same 18\%, to remove the
remaining events of questionable 2-stop pedigree.

The small peak at $x_{track} - x_{tag} \approx 0$ in the simulated background in
Fig.~\ref{fig:tail2} indicates that a small fraction of the punch-through event sample
used in the simulation really corresponds to true 2-stop events that were misidentified by
virtue of BD noise that evaded the noise cuts discussed in the preceding subsection.
Subtracting this small fraction of valid 2-stop events along with the simulated corrupted
events has the effect of reducing the 2-stop tagged neutron yield by $\approx 3\%$ in all
three event streams, with no significant consequence for the absolute $np$ cross sections
extracted from the 2-stop sample.

\subsection{Background subtraction}
\label{backsub}

The background events for this experiment came mostly from np quasifree scattering off
carbon nuclei in the CH$_2$ target.  However, there were also some prominent sources
displaced from the secondary target, including: (1) protons coming directly from the gas
jet production target, or from the exit flange on the Cooler beam $6^\circ$ magnet chamber
(see Fig.~\ref{fig:setup}), that evaded the veto scintillators due either to their
imperfect coverage or electronic inefficiencies; (2) np scattering events induced either
on the downstream scintillator face or the Lucite light guide for the SUV, yielding pulse
heights below that veto detector's threshold; and (3) quasifree np scattering induced on
the vertically narrow (but longitudinally thick) aluminum frame used to support the
secondary target. By frequently interchanging the CH$_2$ target with a graphite target
closely matched in transverse dimensions and in areal density of carbon nuclei, we were
able to subtract the backgrounds from all sources simultaneously. The relative
normalization of the CH$_2$ and C runs was determined from the pd elastic scattering yield
from the GJT, as recorded in event stream 4 \cite{Pet2004}.  The background subtraction
was determined to be sufficiently reliable that we could avoid imposing many kinematic
cuts, with potentially significant systematic ambiguities, to define free np scattering
events.

\begin{figure*}[htp!]
\begin{center}
\includegraphics[scale=0.25]{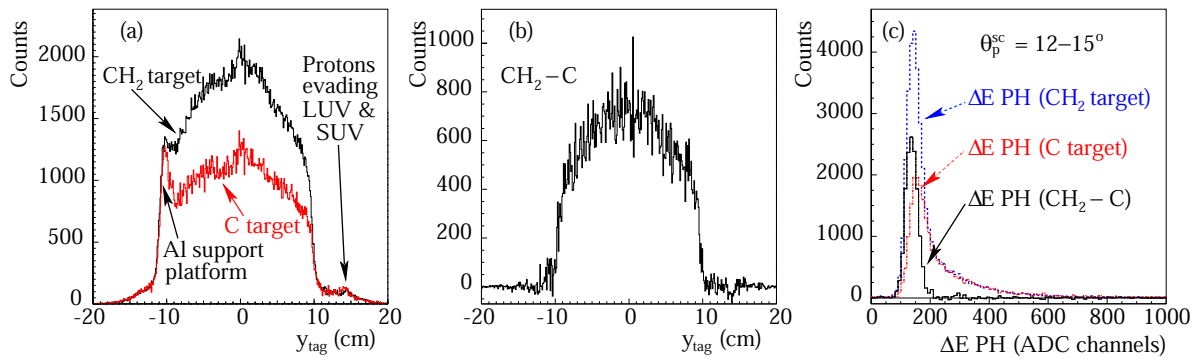}
\caption{\label{fig:bkgrnd}\small Distribution of np scattering event candidates with
respect to $y_{tag}$ and to $\Delta E$ scintillator pulse height for the CH$_{2}$ and C
targets separately, and for the difference between them.}
\end{center}
\end{figure*}

The accuracy of the background subtraction can be judged, for example, from
Fig.~\ref{fig:bkgrnd}, which presents CH$_2$ and C spectra, and their difference, with
respect to $y_{tag}$ (the vertical impact position of the neutron on the secondary target,
as reconstructed from the tagger) and $\Delta E$ scintillator pulse height within a narrow
np scattering angle range.  These two particular variables have been chosen for display in
the figure because the CH$_2$ spectra show prominent background features associated both
with quasifree scattering (the long high pulse height tail in $\Delta E$) and with other
sources (the $y_{tag} = -11$ cm peak from the aluminum support frame). Both sources are
precisely eliminated by the background subtraction.  Indeed, upper limits on the surviving
remnants of these features allow us (as described in Sec.~\ref{backsubunc}) to reduce the
systematic uncertainty for the background subtraction even below the level given by the
precision of the measured C/CH$_2$ target thickness ratio ($\pm 0.6\%$).

\subsection{Free scattering cuts}
\label{freesection}

Software conditions imposed only on event stream 2 to distinguish free-scattering from
background events have the potential to remove free-scattering yield in sometimes subtle
ways.  They were thus used in the analysis only when they could substantially reduce the
statistical uncertainties (\emph{i.e.}, by suppressing background to be subtracted)
without introducing significant systematic uncertainties in correcting for the
free-scattering losses, or when such losses were judged to be inevitable to remove
ambiguities in the analysis. The accuracy of the C background subtraction provided a
reliable method to judge the extent of any free-scattering event removal.

\begin{figure}[htp!]
\begin{center}
\includegraphics[scale=0.4]{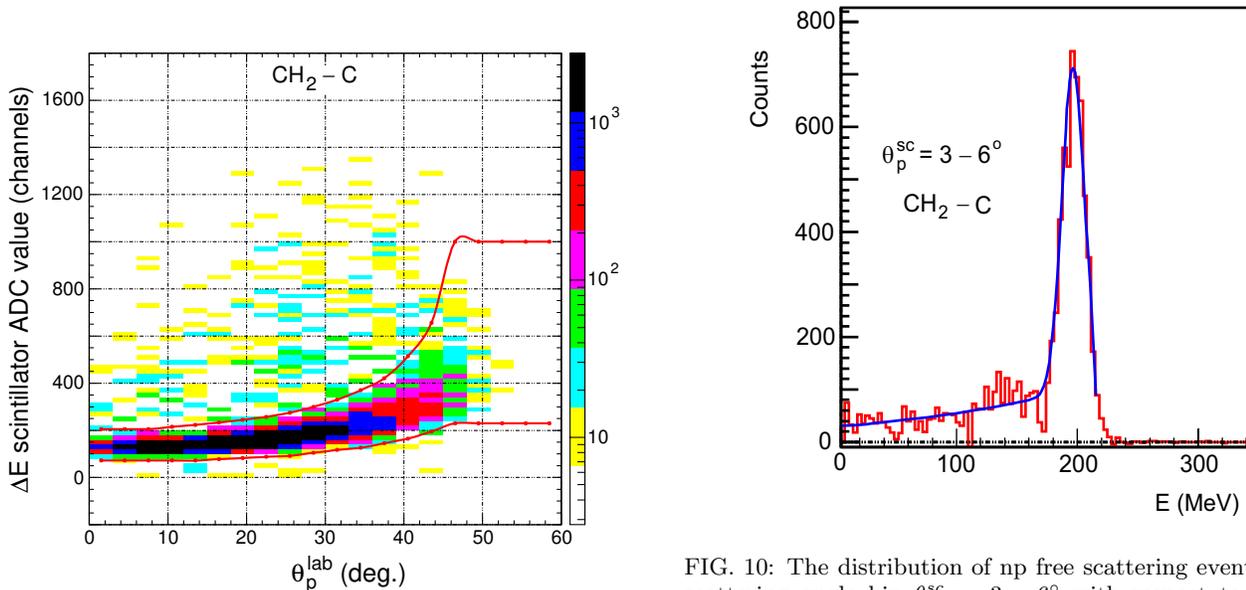}
\caption{\label{fig:syserr_ii.f}\small The distribution of np scattering candidate events
after subtraction of C from CH$_2$ data with respect to $\Delta E$ ADC and proton lab
angle. The red lines show the boundaries of the gate applied to event stream 2 to select
free scattering events.}
\end{center}
\end{figure}

The most effective such cut applied was placed on the correlation of forward proton energy
loss in the $\Delta E$ scintillator with the laboratory angle of the proton trajectory.
The applied two-dimensional software gate is superimposed on the observed distribution of
events following CH$_2$ - C subtraction in Fig.~\ref{fig:syserr_ii.f}. This distribution
reveals that very few free-scattering events were removed by this gate, but it is clear
from the long tail seen in the projected unsubtracted spectrum for one angle bin in
Fig.~\ref{fig:bkgrnd}(c) that a substantial number of quasifree background events, leading
to lower-energy outgoing protons, were successfully removed.

\begin{figure}[htp!]
\begin{center}
\includegraphics[scale=0.5]{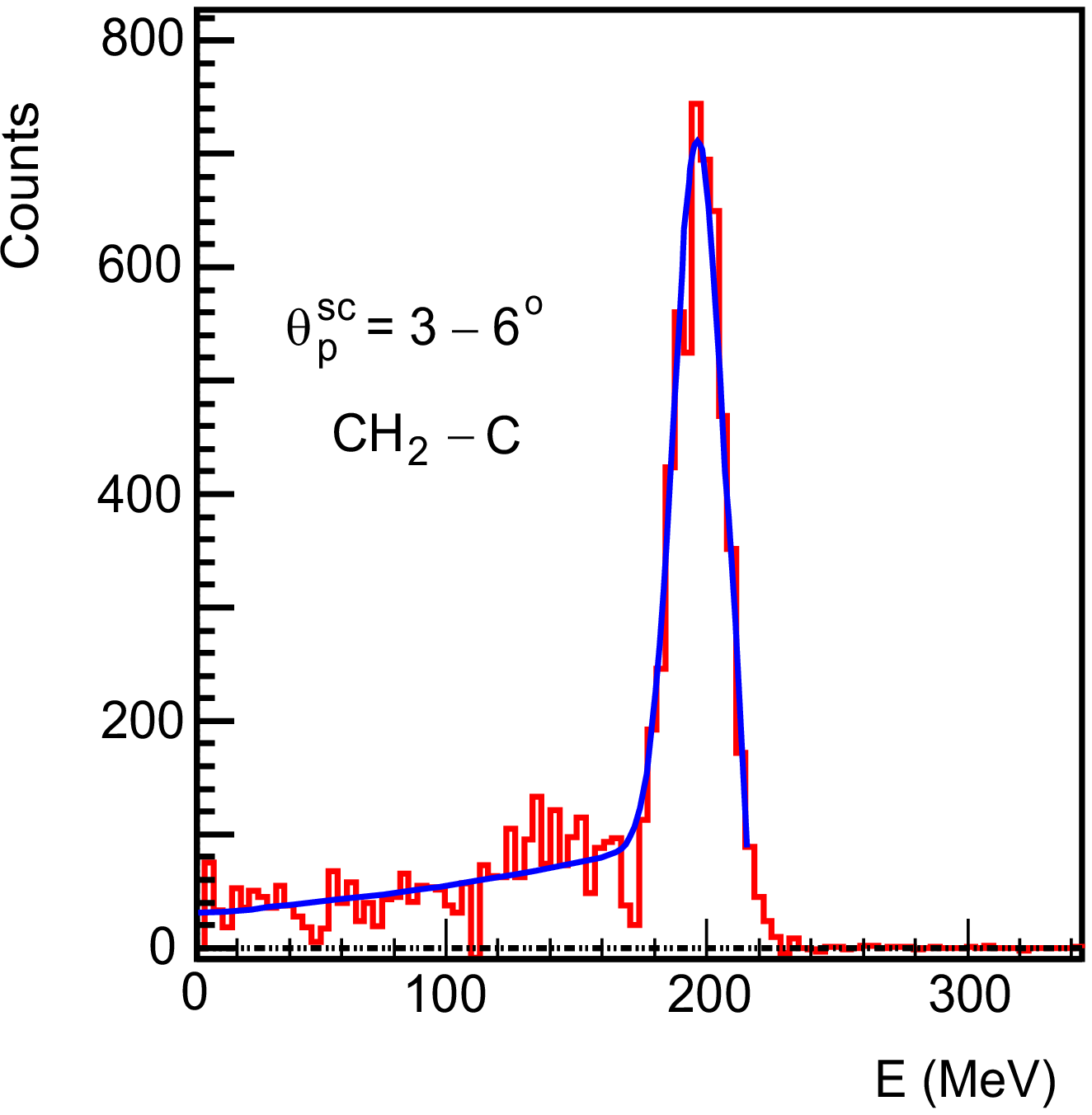}
\caption{\label{fig:syserr_ii.d}\small The distribution of np free scattering events in
the scattering angle bin $\theta^{sc}_p = 3-6^\circ$ with respect to forward proton energy
deposition in the rear hodoscope.  The curve represents a fit with a Gaussian plus
exponential tail.  The tail represents valid free scattering events where the proton
undergoes a nuclear reaction in the stopping scintillator.}
\end{center}
\end{figure}

In contrast, we did not apply a comparable cut on the energy deposition of the forward
proton in the rear hodoscope, where it generally stopped, despite an appreciable
difference in the distributions of hodoscope energy between free and quasifree events.
The reason for avoiding this cut is illustrated in Fig.~\ref{fig:syserr_ii.d}: the
free-scattering spectrum revealed by the C subtraction exhibits a quite substantial
low-energy reaction tail in addition to the well-defined full-energy peak.  An
unacceptably large systematic error would have been introduced by the need to correct for
loss of these reaction tail events, if we had imposed a cut on hodoscope energy to
suppress background.

A benign cut imposed for slight background reduction placed an upper threshold on the
$\chi^2$ value obtained for a linear track fit to the reconstructed MWPC space points. The
CH$_2$ - C subtraction indicates that only $(0.2 \pm 0.1)\%$ of free-scattering events
were removed by this condition.  More serious (6.3\% of total CH$_2$ - C yield), but
unavoidable, losses were introduced by cuts confining the tagging and tracking information
in event stream 2 to agree within $|x_{track} - x_{tag}| \leq 2.5$ cm and $|y_{track} -
y_{tag}| \leq 2.0$ cm.  These limits correspond to $\pm 3\sigma$ of the narrow Gaussian
resolution function that dominates these distributions in the CH$_2$ - C spectra.
Nonetheless, the cut eliminates events in long distribution tails that are affected either
by tagging errors or sequential reactions of the tagged neutron, which introduce serious
ambiguities in interpretation.  This cut, and its consequence for systematic
uncertainties, will be discussed further in Sec.~\ref{seqreac}.

Finally, it is worth mentioning one additional cut that we chose \emph{not} to impose. The
transverse np vertex coordinates are, in fact, determined by the tagging and tracking with
considerably better resolution than implied by the $\sigma \approx 7-8$ mm value mentioned
in the preceding paragraph \cite{Pet2004}.  This latter value is dominated by the
thickness of the secondary target, simply reflecting the uncertainty in precise
longitudinal origin of the np scattering vertex.  Much better information is, in
principle, available by locating the vertex in three dimensions at the point of closest
approach of the neutron trajectory reconstructed from the tagger and the proton trajectory
reconstructed from the MWPC's. Distributions of such reconstructed secondary vertex
coordinates \cite{Pet2004} permit tagged neutron radiography of the background sources
displaced from the CH$_2$ target. However, any cuts to remove such background sources in
this manner would be affected by the strong dependence of the reconstructed vertex
resolution on the proton scattering angle (vertex information clearly deteriorates as the
neutron and proton trajectories approach collinearity).  We decided to rely completely on
the C subtraction to remove such other sources of background, in order to avoid consequent
angle-dependent free-scattering event losses.

\subsection{Acceptance}
\label{accepsection}

The lab-frame proton scattering angle $\theta_p^{sc}$ is determined for each analyzed
event as the opening angle between the neutron trajectory reconstructed from the tagger
and the forward proton trajectory reconstructed from the MWPC's. The geometric acceptance
of the forward detector array for np scattering events is a function of both
$\theta_p^{sc}$ and the coordinates of the scattering vertex at the secondary target.
Because the distribution of scattering vertex coordinates, especially of $x_{tag}$ (see
Fig.~\ref{fig:enxtag}), differed among the three analyzed data subsamples (1-punch, 2-stop
with $E_{p1} \leq 5.0$ MeV, and 2-stop with $E_{p1} > 5.0$ MeV), the acceptance had to be
evaluated separately for each subsample.  This was done by comparing simulated to measured
distributions of events with respect to azimuthal angle $\phi_p^{sc}$ within each
$\theta_p^{sc}$ bin, for each data subsample.

The simulations were constrained to reproduce the measured distributions of the
longitudinal production vertex coordinate of the neutron within the GJT (common to all
three data subsamples) and its transverse coordinates on the secondary target (separately
for each subsample). Within these distributions, coordinates were generated randomly for
each event, as were also $\theta^{sc}_{p}$  (in the range 0--75$^\circ$) and $\phi_p^{sc}$
(over the full azimuthal range). Generated outgoing proton trajectories were then accepted
if they would yield signals above the hodoscope pulse height threshold (required in
trigger) and in all three MWPC's (required in the data analysis). Forward detector
location parameters were tuned slightly from their measured values to optimize the fit of
the simulated to the measured $\phi_p^{sc}$ distributions for all $\theta_p^{sc}$ bins and
for 1-punch and 2-stop samples simultaneously.

\begin{figure}[htb!]
\begin{center}
\includegraphics[scale=0.2]{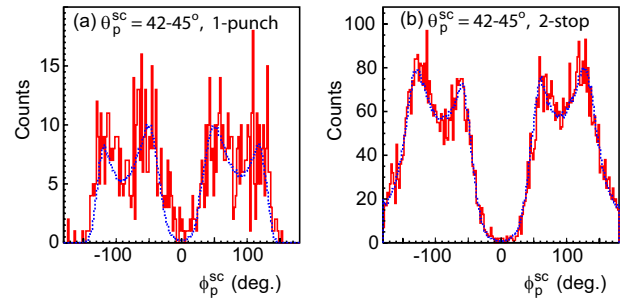}
\caption{\label{fig:acceptance}\small Comparison of the measured (solid line with sizable
statistical fluctuations) and simulated (dotted line) distributions of free (CH$_2$-C) np
scattering events in the angle bin $\theta_p^{sc} = 42-45^\circ$ with respect to proton
azimuthal scattering angle $\phi_p^{sc}$, for the 1-punch (a) and 2-stop (b) data samples.
Forward detector geometry parameters, plus a single overall normalization parameter per
angle bin and data sample, have been adjusted to optimize the fit simultaneously for all
angle bins and both data samples.}
\end{center}
\end{figure}

\begin{figure}[htb!]
\begin{center}
\includegraphics[scale=0.4]{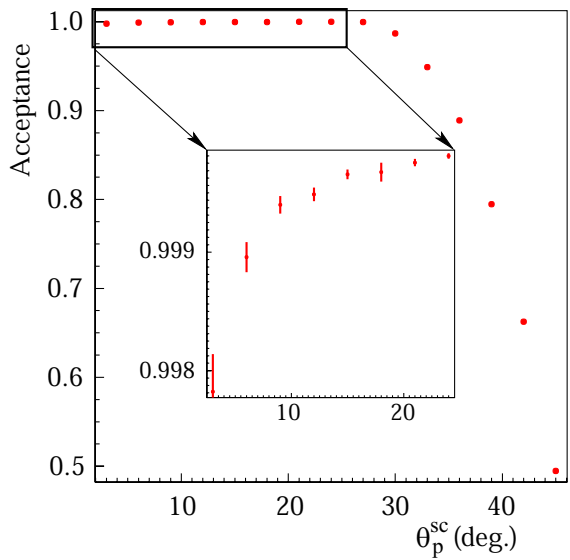}
\caption{\label{fig:acceptance2}\small The simulated acceptance of the 1-punch data
sample.  The inset shows a greatly magnified vertical scale for the most forward proton
scattering angles.}
\end{center}
\end{figure}

The high quality of the fits obtained is illustrated in Fig.~\ref{fig:acceptance} for the
1-punch (a) and 2-stop (b, summed over all $E_{p1}$) samples for a single large angle bin,
$\theta^{sc}_p$=42-45$^\circ$, where the observed azimuthal distributions display
considerable structure. The structure reflects the rectangular shape of the hodoscope and
large MWPC, projected onto $\theta - \phi$ space: \emph{e.g.}, the four peaks observed
correspond to the four detector corners.  The small changes in distribution between the
1-punch and 2-stop samples -- \emph{e.g.}, in the relative heights of the peaks and in the
extent of the dips near $\phi_p^{sc} = 0^\circ$ (beam left side) and 180$^\circ$ (beam
right) -- arise from the shift in $x_{tag}$ profiles seen in Fig.~\ref{fig:enxtag}. These
features are all reproduced very well by the simulations.  For $\theta_p^{sc} \leq
24^\circ$, the measured and simulated $\phi$ distributions are essentially uniform over
2$\pi$, indicating full acceptance. Figure~\ref{fig:acceptance2} shows the simulated
acceptance for the 1-punch data sample as a function of $\theta_p^{sc}$.  The 0.2\%
shortfall from full acceptance near 0$^\circ$ reflects protons incident normally on the
small cracks between adjacent hodoscope elements.  Results presented in the next section
are limited to the angle range for which the acceptance is at least 50\%; at larger proton
angles the uncertainty in acceptance grows rapidly.

\section{RESULTS}
\label{finalresults}

The absolute differential cross section for np back-scattering was extracted independently
for three data samples -- 1-punch, 2-stop with $E_{p1} \leq$5 MeV, and 2-stop with
$E_{p1}>$5 MeV -- from the yields of event streams 1, 2 and 3 as follows:

\begin{equation}
\label{diffcrosssec}
(\frac{d \sigma}{d \Omega})_{lab}=\frac{N_2(\theta_p^{sc}) \prod
c_i}{(N_1 + N_2 + N_3) t_H |d \cos(\theta_{p}^{sc})|
a_{\phi}(\theta_p^{sc})},
\end{equation}

\noindent where $N_2(\theta_p^{sc})$ represents the number of free-scattering events from
stream 2 within a given reconstructed proton angle bin, surviving all relevant cuts and
background subtractions; $N_1, ~N_2$ and $N_3$ in the denominator represent analogous
tagged-neutron yields from the mutually exclusive event streams 1 (corrected for
prescaling), 2 (angle-integrated) and 3; the $c_i$ represent small corrections, summarized
in Table~\ref{tab:syserr} with details in section \ref{syserrors}, for various
inefficiencies, tagged neutron losses or backgrounds, and software cut and dead time
differences among event streams; $t_H = (1.988 \pm 0.008) \times 10^{23}$ H atoms/cm$^2$
for the CH$_2$ target; and $a_{\phi}$ is the azimuthal acceptance determined from
simulations for the given angle bin.  The data were analyzed in 1 MeV wide slices of
reconstructed neutron energy from 185 to 197 MeV and an effective cross section extracted
at the mean neutron energy of 194.0$\pm$0.15 MeV.  For this purpose, a small (always
$<1\%$) cross section correction was made for the deviation of each analyzed slice from
the mean energy, using the theoretical energy- and angle-dependence calculated with the
Nijmegen PWA93 solution \cite{Sto1993}.

\begin{table}[htp!]
\caption{\label{tab:syserr} Correction factors and systematic
uncertainties in correction factors for the np cross sections.}
\begin{ruledtabular}
\begin{tabular}{lll}
Source & Correction Factor (c$_i$) & Uncertainty in c$_i$ \\
\hline\hline
Accid. tagger coinc.                  & 1.0003               & $<$ $\pm$ 0.001 \\
Non-D$_{2}$ tagger                  & 1.0067 (2-stop);     & $\pm$ 0.002 \\
~~background                        &  1.0044 (1-punch)    & \\
n pos'n unc. on CH$_{2}$             & 1.0000               & $\pm$ 0.001 \\
n atten'n before CH$_{2}$                  & 1.005                & $\pm$ 0.0025 \\
Sequential react'ns                       & 1.063    & $\pm$ 0.010 \\
~~\& $x_{tag}$(n) errors                  & & \\
C bkgd. subtraction                       & 1.0000               & $\pm$ 0.004 \\
H in C target                             & 1.000                & $\pm$ 0.004 \\
Corrupted event                     & 1.000                & $< \pm 0.001$ \\
~~subtraction  & & \\
Software cut losses                       & 1.010                & $\pm$ 0.005 \\
Reaction tail losses                           & 1.004                & $\pm$ 0.002 \\
Neutron polarization                    & Angle-dependent:    & $\pm$ 0.001 \\
~~effects                               & $<1.0012$ (1-punch)    & \\
                                        & $>0.9986$ (2-stop)     & \\
CH$_{2}$ tgt. thickness                      & 1.0000               & $\pm$ 0.004 \\
np scattering                  & 1.0000 & $\leq\pm 0.001$ ($>$120$^\circ$)  \\
~~acceptance                         &  & $\rightarrow \pm 0.017$ (90$^\circ$)\\
MWPC inefficiency                              & 1.017                & $\pm$0.002 \\
Trigger inefficiency                           & 1.002 + 0.008 $\times$ & $\pm$ [0.001 + 0.004 \\
                                               & cos$^{2}$($\theta_{p}^{LAB}$)
                                               & $\times$ cos$^{2}$($\theta_{p}^{LAB}$)]\\
Dead time diffs.                          & 0.991                & $\pm$ 0.005 \\
Scattering angle                        & 1.000                & angle-dependent, \\
~~errors                                &                      & ~~$\leq \pm$0.004 \\
\hline {\bf Net, typical}               & $\approx 1.10$       & $\approx \pm 0.016$ \\
\end{tabular}
\end{ruledtabular}
\end{table}

\begin{figure}[htp!]
\includegraphics[scale=0.4]{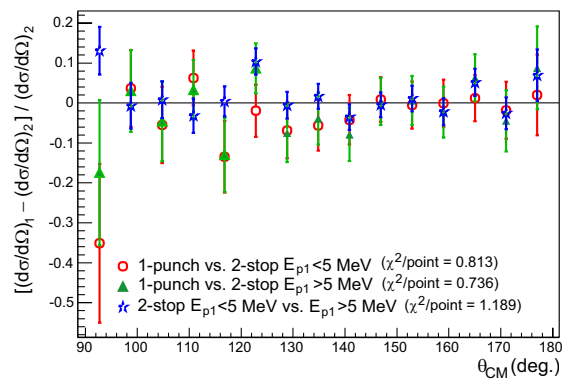}
\caption{\label{fig:com1pn2s} The fractional differences between the absolute differential
cross sections extracted for different analyzed data subsamples. The plotted error bars
take into account only the independent statistical (including those from background
subtractions) uncertainties for the three samples.  Slightly different correction factors
$c_i$ were applied to the cross sections for different samples, as indicated in
Table~\ref{tab:syserr}, before the comparison was made.}
\end{figure}

The cross sections for the three data subsamples, with their independently determined
absolute scales, are mutually consistent in both magnitude and angular shape, within
statistical uncertainties, as revealed by the comparisons in Fig.~\ref{fig:com1pn2s}. The
figure shows the relative difference, $((\frac{d\sigma}{d\Omega})_{sample A} -
(\frac{d\sigma}{d\Omega})_{sample B})/ (\frac{d\sigma}{d\Omega})_{sample B}$, between
pairs of cross sections for the three data samples.  The reduced $\chi^2$ value for the
comparison of each pair of samples is indicated in the legend of Fig.~\ref{fig:com1pn2s}.
This comparison supports the reliability of the experiment and analysis, because these
samples come from complementary regions of the tagged beam spatial and energy profiles
(see Fig.~\ref{fig:enxtag}) and are subject to somewhat different systematic error
concerns. We view the agreement in absolute cross section scale as particularly
significant demonstrations of the accuracy of the neutron profiles reconstructed from
tagging and of the subtraction procedure applied to remove corrupted events from the
2-stop sample (see Sec.~III.B). Cross sections extracted for different time periods within
the production runs, and with different sets of cuts, are also consistent within
uncertainties.

\begin{figure}[htp!]
\includegraphics[scale=0.45, angle=-90]{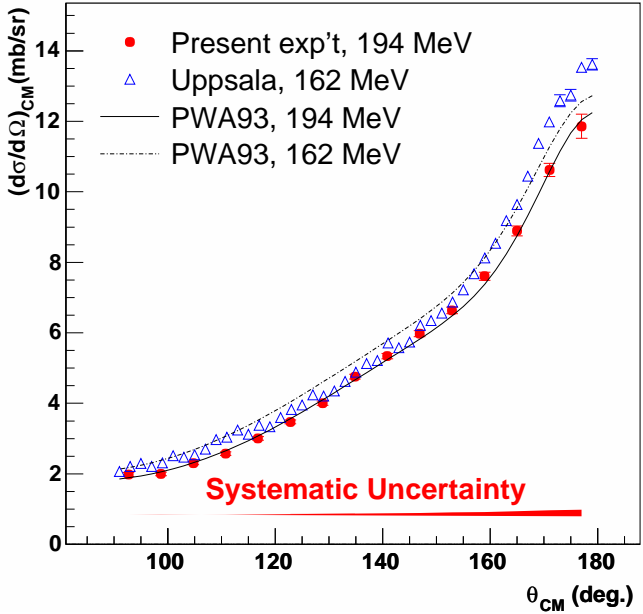}
\caption{\label{fig:avexs} Absolute differential cross section from the present
experiment, compared with data from Ref.~\cite{Rah1998} and with PWA calculations at two
relevant energies.  The error bars on the present results are statistical (including
background subtraction), while the shaded band represents all systematic uncertainties,
including those in the overall normalization.}
\end{figure}

The results, averaged over all three data samples, are compared in Fig.~\ref{fig:avexs}
with previous experimental results at 162 MeV \cite{Rah1998} and with the Nijmegen partial
wave analysis (PWA93) at the two relevant energies \cite{nij}.  The measured points are
plotted at the yield-weighted centroid angle of each analyzed bin.  The comparison of the
present results with previous experiments and with partial wave analyses will be discussed
in detail in Sec. VI, after first describing the nature and evaluation procedure for each
of the systematic uncertainties included in Table~\ref{tab:syserr}.

\section{SYSTEMATIC ERRORS}
\label{syserrors}

Most of the individual correction factors $c_i$ applied to the extracted cross sections,
and their associated systematic uncertainties listed in Table~\ref{tab:syserr}, have been
evaluated via complementary analyses of the data.  In this section we briefly describe the
procedures used and error estimates for each, being careful to distinguish uncertainties
that affect only the overall cross section normalization from those with appreciable
angle-dependence.  In the latter cases, we also characterize the degree of correlation
among the uncertainties at different angles, to facilitate inclusion of the uncorrelated
systematic errors in PWA's including the present data. For purposes of logical flow, we
organize the discussion into four categories: (1) tagged neutron flux uncertainties; (2)
np backscattering yield uncertainties; (3) target thickness, acceptance and efficiency
uncertainties; and (4) errors in determining the kinematic variables.

\subsection{Tagged neutron flux uncertainties}

The sources below contribute to uncertainties in extracting the angle-integrated yields
N$_{1,2,3}$ in Eq.~(\ref{diffcrosssec}), dominated by the non-interacting tagged neutrons
in event stream 1.  All of the issues discussed in this subsection give rise to overall
(angle-independent)
normalization errors in the differential cross sections.\\

\subsubsection{Accidental tagger coincidences}

\begin{figure}[htp!]
\includegraphics[scale=0.45]{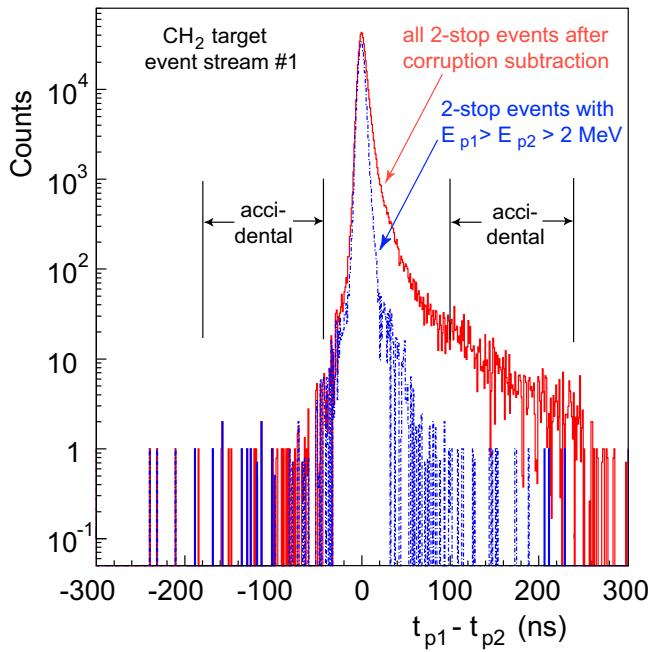}
\caption{\label{fig:tp1tp2} The distribution of DSSD arrival time difference between the
two recoil protons for tagged neutron events, shown for all 2-stop events (following
subtraction of the corrupted events as per Sec.~\ref{corrupt}) and for 2-stop events where
the lower of the two DSSD energy depositions ($E_{p2}$) exceeds 2.0 MeV. The vertical
lines indicate the prompt coincidence gate (central region between the inner two lines)
imposed in the analysis, plus two displaced gates used to assess accidental coincidence
background in event stream 1. The long tail seen for events with $E_{p2} \leq 2.0$ MeV,
arising from detector noise and imperfect software corrections for time walk near the
front-end discriminator threshold, leads to an overestimate of the accidental coincidence
yield, but the correction and uncertainty still remain quite small.}
\end{figure}

Accidental coincidences between two uncorrelated particles detected in the tagger
contribute slightly to the apparent tagged neutron flux on the secondary target, leading
to an underestimate of the cross section. The accepted events in all three event streams
passed a cut on the time difference $\Delta t=(t_{p1}-t_{p2})$ between the two tagger
hits, as indicated in Fig.~\ref{fig:tp1tp2}.  The correction factor was determined from
the ratio of events in stream 1 that passed all other cuts defining the tagged neutron
beam but fell within one of two displaced time windows, $\mid \Delta t+110~{\rm ns} \mid
\le 70$ ns and $\mid \Delta t-170~{\rm ns} \mid \le 70$ ns, to the yield in the prompt
coincidence window $\mid \Delta t-30~{\rm ns} \mid \le 70$ ns.  The resulting correction
factor is $c_1 = 1.0003$, with an uncertainty $< \pm 0.001$, showing that accidental
coincidences were a minor issue for the experiment.

\subsubsection{Tagger background from non-D$_2$ sources}

Additional possible background contributions to the tagged neutron flux could arise from
real (correlated) two-particle coincidences in the tagger, generated by proton beam
interactions with nuclei heavier than deuterium in material displaced from the GJT. This
possibility was checked via runs where H$_2$ was substituted for the D$_2$ in the GJT, to
induce similar beam ``heating" without any real possibility of tagged neutron production
(since the proton beam energy was far below pion production threshold for the p+p system).
A correction factor $c_2 = 1.0044\pm 0.002 ~(1.0067\pm 0.002)$ for 1-punch (2-stop) events
was determined from the ratio of accidental-subtracted tags satisfying the tagged neutron
conditions with the H$_2$ \emph{vs.} D$_2$ production targets. The statistical
uncertainties in these ratios were considerably smaller than $\pm 0.002$; the quoted
uncertainty is intended to allow for the possibility of slight systematic differences in
beam heating, hence in the rate of interactions with displaced material, between the two
GJT gases.

\subsubsection{Impact position uncertainty on the CH$_{2}$ target}
\label{posunc}

This is the first of several error sources we consider that arise when a properly tagged
neutron does not reach the CH$_2$ (or C) secondary target, or reaches it at a
significantly different position than expected from the tagging.  Because of the finite
(several mm) impact position resolution from the tagger, some tagged neutrons predicted to
hit the secondary target may actually miss it, while some predicted to miss the target may
hit it. Especially near the target edges, where the yield of np scattering events drops
rapidly and nonlinearly as a function of impact position, this resolution smearing can
affect the extracted cross sections. In practice, however, we observe no statistically
significant difference in cross section normalization between the independent event
samples from the bulk of the target ($\mid x_{tag} \mid \leq 9.0$ cm and $\mid y_{tag}
\mid \leq 9.0$ cm) and from a 5.0-mm wide strip ($9.0 < \mid x_{tag} \mid \leq 9.5$ cm or
$9.0 < \mid y_{tag} \mid \leq 9.5$ cm) surrounding this core. From this comparison and the
fraction of all events arising near the target edges, we infer a correction factor $c_3 =
1.000 \pm 0.001$.

\subsubsection{Neutron attenuation before the CH$_{2}$ target}
\label{nattenbefch2}

Some tagged neutrons fail to hit the secondary target, leading to an underestimate of the
extracted np cross section, as a result of interactions they undergo upstream of that
target. Approximately 3.5\% of 200 MeV neutrons will undergo an inelastic reaction of some
sort in the upstream material \cite{NASA1997}, which is dominated by the 0.29 cm thick
stainless steel vacuum window at the exit of the Cooler's 6$^\circ$ magnet vacuum chamber,
the 0.64 cm thick LUV plastic scintillator, and the 0.64 cm thick SUV plastic scintillator
(the first two of these traversed at an incidence angle $\approx 14^\circ$).  However,
many of these ``pre-scattering" neutrons give rise to charged products that get vetoed by
LUV or SUV (and hence do not contribute to the tagged flux) or are removed by the C
subtraction as apparent np scattering events from an upstream source.  Others yield an
energetic neutron or proton, not strongly deflected from the original tagged neutron
trajectory, that still strikes the nearby secondary target with a chance to induce a
tertiary interaction there, and so might still be considered as part of the incident flux.

\begin{figure}[htp!]
\includegraphics[scale=0.26]{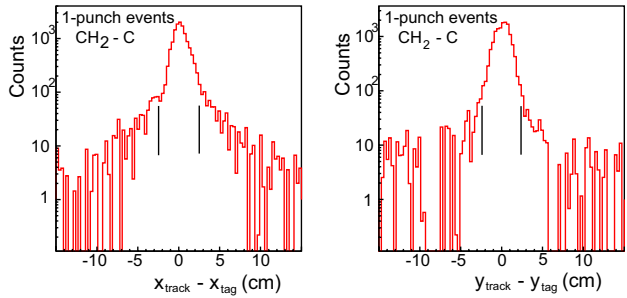}
\caption{\label{fig:dxny_1p} The difference distributions in $x$ (left) and $y$ (right)
coordinates determined from tracking of the forward proton \emph{vs.} reconstruction of
the tagged neutron, for the 1-punch data sample after CH$_2$ - C subtraction. The vertical
lines indicate the location of software gates used to remove events that may be
complicated by sequential reactions or tagging errors.}
\end{figure}

We may judge the rate of such tertiary interactions from events where a forward proton
emerges from the secondary target at a transverse location ($x_{track}, y_{track}$)
substantially different (by much more than the tagging position resolution effect
considered in Sec.~\ref{posunc}) from the predicted impact position of the tagged neutron
($x_{tag}, y_{tag}$). Such tertiary interactions introduce their own problems in the
analysis, to be addressed separately in the next subsection.  Here, we study them in order
to place limits on the probability of larger-angle upstream scattering, which yields no
chance of a tertiary scattering.  In Fig.~\ref{fig:dxny_1p}, we show the difference
spectra for $x_{track} - x_{tag}$ and $y_{track} - y_{tag}$ for 1-punch events (to
eliminate ambiguities from the corrupted 2-stop events seen in Fig.~\ref{fig:tail1}) after
CH$_2$ - C subtraction (to eliminate ambiguities from np scattering induced on material
displaced from the secondary target).  The narrow Gaussian resolution peaks sit atop a
broad background that has important contributions from these tertiary interactions.

By fitting the background in Fig.~\ref{fig:dxny_1p} with a broad Gaussian, and assuming
that the probability of initiating a scattering in the secondary target is roughly the
same for the pre-scattered neutrons as for the bulk of the tagged neutrons, we estimate
that 0.5\% of tagged neutrons may be pre-scattered through a sufficiently large angle to
cause a transverse displacement greater than 10 cm (\emph{i.e.}, half the target width or
height) on the secondary target. We use this estimate to infer a correction factor $c_4 =
1.005 \pm 0.0025$ for tagged neutron pre-scattering flux losses before the CH$_2$ target.
The $\pm 50\%$ uncertainty we assign to $(c_4 - 1)$ is intended to account for
non-prescattering origins of the background in Fig.~\ref{fig:dxny_1p} (\emph{e.g.},
tagging errors or sequential reactions \emph{within} the CH$_2$ or tertiary scattering of
a forward proton in material \emph{following} the CH$_2$ target) and for possible upstream
neutron interactions that elude the above analysis.

\subsubsection{Sequential reactions in the secondary target or upstream material}
\label{seqreac}

Here we deal explicitly with the events contributing to the broad backgrounds in
Fig.~\ref{fig:dxny_1p} and in the analogous distributions for 2-stop events. From the
behavior of these events and the differential cross sections we extract specifically from
them, we judge them to correspond primarily to valid free np scattering in CH$_2$ either
following an earlier interaction of the tagged neutron or preceding a later interaction of
the forward proton. The two interactions will in some cases both have taken place within
the CH$_2$ target. Some of the background may also arise from tagging errors associated
with less than complete energy collection for the recoil protons in the tagger. Regardless
of their detailed source, such events are distorted because we mismeasure the scattering
angle and possibly the incident neutron energy for the free np scattering.

The least biased way to handle these events is to eliminate them from the analyzed sample.
This is simple enough to do for each angle bin in event stream 2, via the software cuts
requiring $|x_{track} - x_{tag}| \leq 3\sigma_x$ and $|y_{track} - y_{tag}| \leq
3\sigma_y$, where $\sigma_x \approx 0.8$ cm and $\sigma_y \approx 0.7$ cm are the widths
of the narrow Gaussian peaks in Fig.~\ref{fig:dxny_1p}.  However, we have no access to
analogous cuts for event streams 1 and 3, where there is no MWPC information.  Hence, we
must correct the extracted cross section for our inclusion of tagged neutron flux that is
associated with such eliminated sequential reaction events. The correction factor $c_5 =
1.063 \pm 0.010$ is the largest single correction and systematic uncertainty we apply. The
value assumes that the 6.3\% of event stream 2 (CH$_2$ - C) yield (averaged over 1-punch
and 2-stop samples and over scattering angle) removed by these sequential reaction cuts
arises from 6.3\% of the tagged neutron flux. (The actual correction factors applied
differ for the three data samples, reflecting differences in the fraction of events
removed by these cuts.)  The uncertainty allows for errors in this assumption, for
example, because it neglects the energy-dependence of the np scattering probability in the
CH$_2$ target.

\subsection{Uncertainties in absolute np backscattering yields}

Analysis issues in the extraction of the free-scattering yield $N_2(\theta)$ needed in
Eq.~\ref{diffcrosssec} can lead, in principle, to angle-dependent errors.  We thus specify
for each case below whether the estimated uncertainty should be considered as
angle-dependent and as uncorrelated from angle bin to bin.

\subsubsection{Uncertainties in background subtraction via the C target}
\label{backsubunc}

As described in Sec.~\ref{backsub}, we relied heavily on the CH$_2$ - C subtraction to
remove simultaneously backgrounds due to quasifree scattering from carbon nuclei in the
secondary target and to reactions induced on displaced sources.  The precision of the
subtraction depends on that of our knowledge of the relative target thicknesses and
integrated neutron flux exposures for the CH$_2$ \emph{vs.} C runs, and on the stability
of beam conditions between the two sets of runs. The relative normalization, taken from
cleanly (kinematically) identified pd scattering yields measured simultaneously, is
determined with quite high statistical precision but could, in principle, deviate
systematically from the more relevant ratio of tagged neutron yields. The overall
precision of the relative normalization was judged from the extent to which scattering
events from the aluminum target platform (see Fig.~\ref{fig:bkgrnd}(a)) were successfully
removed by the C subtraction. We concentrate first on this background source because its
yield is not sensitive to the CH$_2$/C target thickness ratio.

\begin{figure}[htp!]
\begin{center}
\includegraphics[scale=0.44]{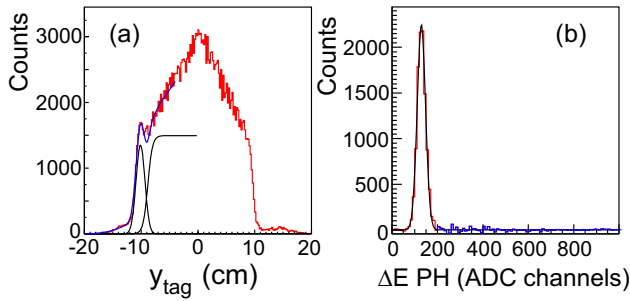}
\caption{\label{fig:bckgrnd_ii.a}\small The reconstructed (a) $y_{tag}$ spectrum for the
CH$_{2}$ target before background subtraction, and (b) $\Delta E$ pulse-height spectrum
for CH$_2$-C for the $\theta_p^{sc}$ bin 3-6$^\circ$.  The superimposed curves in both
frames represent fits used in estimating background subtraction accuracy.}
\end{center}
\end{figure}

The reconstructed $y_{tag}$ distributions in the vicinity of the aluminum platform peak,
for both CH$_2$ and C targets (see Fig.~\ref{fig:bkgrnd}), could be well reproduced by the
sum of a Fermi distribution and a polynomial, to represent the bottom target edge, and a
Gaussian to represent the aluminum peak.  The fit for the CH$_2$ is shown in
Fig.~\ref{fig:bckgrnd_ii.a}(a). An analogous fit was then carried out for the C-subtracted
spectrum in Fig.~\ref{fig:bkgrnd}(b), fixing the positions and widths of the Gaussian and
Fermi-function contributions to their common values for CH$_2$ and C. The ratio of events
in the Gaussian peak after subtraction to that before subtraction is $(1.9 \pm 0.54)\times
10^{-3}$, providing one measure of the accuracy of the background subtraction.

An independent measure was provided for each np scattering angle bin by the fraction of
high energy-loss tail events that survive the subtraction in $\Delta E$ pulse-height
spectra (see Fig.~\ref{fig:bkgrnd}(c)). The tail events were integrated by summing all
counts at $\Delta E$ values more than 4$\sigma$ above the center of a Gaussian fitted to
the free-scattering peak, as illustrated in Fig.~\ref{fig:bckgrnd_ii.a}(b).  For the three
largest $\theta_p^{sc}$ bins studied, this approach breaks down because the
free-scattering $\Delta E$ peak develops a substantial Landau tail and is no longer well
reproduced by a Gaussian shape.  But for all (12) smaller-angle bins, the ratio of tail
events after to before C subtraction fluctuates about zero, with a weighted average over
angle bins of $(2.97\pm0.24)\times10^{-3}$.  This measure is sensitive to the target
thickness ratio as well as to the relative flux normalization for CH$_2$/C.  The two
measures combined do not give compelling evidence of a need for any correction, and are
conservatively summarized by associating with the C subtraction an angle-independent
correction factor $c_6 = 1.000 \pm 0.004$.

The uncertainty estimated in this way also subsumes two other potential sources of
systematic error.  One is accidental coincidences between a real tagged neutron and an
uncorrelated forward-going proton emerging from the GJT or the Cooler beam pipe (the most
abundant sources of protons).  To the extent that such coincidences passed all our cuts,
they might have contributed to $N_2 (\theta)$ for the CH$_2$ target.  However, since these
accidentals are independent of the presence or nature of the secondary target, they would
be subtracted via the C target measurements.  The second effect concerns possible proton
attenuation before the hodoscope, which is required as part of the event stream 2 hardware
trigger.  The dominant material between secondary target and hodoscope that might have
served as a source of such proton losses is the $\Delta E$ scintillator, where tertiary
interactions should cause abnormal energy loss.  Since the $c_6$ uncertainty estimate
includes allowance for such abnormal pulse heights in carbon-subtracted $\Delta E$
spectra, it should also include such proton attenuation effects.

\begin{figure}[htb]
\begin{center}
\includegraphics[scale=0.45]{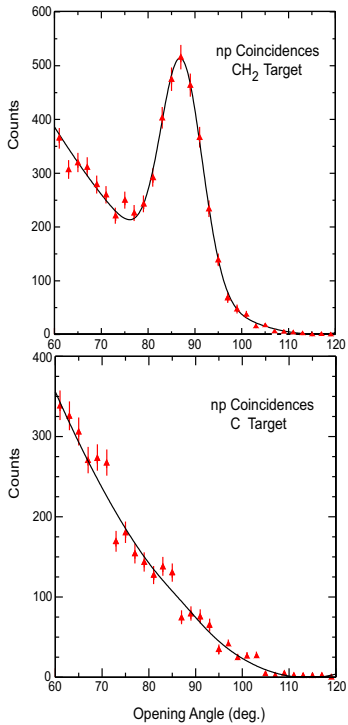}
\caption{\label{fig:freepeak}\small np scattering opening angle spectra reconstructed for
events where a large-angle proton fires at least the first two MWPCs and a forward neutron
appears to fire the rear scintillator hodoscope.  For such coincidence events, a clear
free-scattering peak is seen with the CH$_2$ production target (upper frame), while no
hint of such a peak is seen for the C target (lower).  The solid curves are fits with a
Gaussian peak superimposed on a quadratic background.  The distribution shape for C
reflects the quasifree np scattering opening-angle spectrum convoluted with the
coincidence acceptance of the forward detector array.}
\end{center}
\end{figure}

One further potential complication with the background subtraction could have arisen if
there had been any appreciable hydrogen buildup on the graphite target used for the
subtraction, a possibility limited by the hydrophobic nature of graphite.  In this
circumstance we would subtract some small fraction of the valid free-scattering events,
and thus would introduce an effective overall normalization error in the hydrogen target
thickness $t_H$ used in Eq.~\ref{diffcrosssec}. To estimate this effect we considered np
scattering events forward of $\theta_{c.m.} = 90^\circ$, where event stream 2 contained
some coincidence events, with a forward-scattered neutron detected in the hodoscope and
the larger-angle proton detected in the $\Delta E$ scintillator and (at least) the front
two MWPCs.  The angle of the proton was determined from MWPC ray-tracing, while that of
the neutron was deduced from the hodoscope elements fired and from the position inferred
from the time difference between hodoscope phototubes mounted at the two ends of each
element \cite{Pet2004}. The opening angle spectrum reconstructed for such np coincidence
events exhibited a clear free-scattering kinematic peak for the CH$_2$ target, but only
the Fermi-smeared and acceptance-limited angular correlation characteristic of quasifree
scattering for the C target (see Fig.~\ref{fig:freepeak}). Figure \ref{fig:freepeak}
includes fits to the distributions for both targets based on the sum of a quadratic
background and a Gaussian free-scattering peak, with the peak location and width fixed for
the C target to the values determined from CH$_2$.  The fit for C is statistically
consistent with no hydrogen content in the graphite target, with a 1$\sigma$ limit on the
hydrogen thickness of 0.4\% that of the hydrogen in CH$_2$.  We thus apply a cross section
correction factor $c_7 = 1.000 \pm 0.004$ for hydrogen in the C target.

\subsubsection{Uncertainty in subtraction of corrupted events}
\label{corsuberror}

For the 2-stop event stream, we followed the procedure described in Sec.~\ref{corrupt} to
subtract the punch-through events that had been corrupted by the electronic loss of
backing detector pulse height information.  There is no evidence for any systematic
deviation in distribution shapes between the corrupted sample and our simulation of this
sample using valid recorded punch-through events.  Thus, the only uncertainty we consider
is that in the normalization of the simulated sample to the corrupted events in stream 2.
The normalization factors were determined from fits for the subsample of corrupted events
that had valid backing detector timing signals, and the uncertainty in these normalization
factors was then deduced from the change in normalization factor that caused an increase
of unity in the overall $\chi^2$ value for the fit.  The effect of this normalization
uncertainty on the extracted 2-stop cross sections was typically $\approx 0.01\%$, and is
negligible in comparison with other systematic errors. Hence, we assign a correction
factor $c_8 = 1.000$ with uncertainty $< \pm 0.001$ to the subtraction of corrupted
events.

\subsubsection{Losses via software cuts}

The efficiency of software cuts applied to event stream 2, but not to streams 1 and 3, was
judged by comparing the ratio of cross sections obtained, after CH$_2$ - C subtraction,
for all events failing \emph{vs.} satisfying a given cut.  The most important of these
cuts were on $\Delta E(\theta_p^{sc})$ (see Fig.~\ref{fig:syserr_ii.f}) and on $x_{track}
- x_{tag}$ and $y_{track} - y_{tag}$.  The latter cuts were already dealt with in
Sec.~\ref{seqreac}. (Another cut, on the quality of track fits, is treated together with
wire chamber inefficiencies below.) The $\Delta E$ cut limits were somewhat tighter than
the 4$\sigma$ allowance used in estimating background subtraction accuracy (see
Sec.~\ref{backsubunc}). We found the ratio of background-subtracted events
failing/satisfying the $\Delta E$ cut to be 1.0\%, averaged over all event streams and
angles.  There is no evidence for any significant angle-dependence in this loss, but there
are strong enough fluctuations in the losses from angle to angle or event stream to event
stream that we assign a $\pm 50\%$ uncertainty to the losses. We thus apply a
corresponding, angle-independent correction factor $c_9 = 1.010 \pm 0.005$.  With this
systematic uncertainty, application of the $\Delta E$ cut still reduced the overall cross
section error bars slightly because the quasifree background to be subtracted decreased
significantly.

\subsubsection{Reaction tail losses beneath the hodoscope energy threshold}
\label{reactiontaillosses}

Since we did not use any software cuts on energy deposition in the rear hodoscope, we
avoided the large corrections that would have been needed to account for protons lost to
nuclear reactions in this hodoscope (see Fig.~\ref{fig:syserr_ii.d}).  However, if the
reaction is sufficiently severe that the deposited energy falls below the hodoscope
hardware threshold, then the event will have been lost in hardware to a trigger
inefficiency. To estimate these potential losses, we fit the hodoscope energy spectra
after CH$_2$ - C subtraction to the sum of a Gaussian and an exponential (reaction) tail,
as shown in Fig.~\ref{fig:syserr_ii.d}.  The tail was extrapolated to zero energy
deposition, and the ratio of yields below to above threshold (typically set at 5--10 MeV
proton energy) was thereby estimated.  The loss below threshold was found to be quite
consistent with 0.4\% for each scattering angle bin, so that we again have applied an
angle-independent correction factor $c_{10} = 1.004 \pm 0.002$.

\subsubsection{Neutron polarization effects}

While the stored proton beam in the Cooler was unpolarized for this experiment, the
neutron production reaction selected neutrons scattered to one side of the beam (beam
right) at about $14^\circ$ in the laboratory frame. At this angle, the D(p,n) charge
exchange reaction that dominates our tagged beam production has a small polarization, so
that the beam neutrons would have been slightly polarized vertically (perpendicular to the
horizontal production plane). The magnitude of this effect is P$_n^{prod}\approx-0.1$,
where the minus sign indicates that for neutron production to the right of the cooled
proton beam, the neutron spin points preferentially downward at the secondary target. The
tagged neutron polarization can then give rise to a left-right asymmetry in np scattering
events:
\begin{equation}
\label{syserr_ii.e_1}
    \varepsilon_{np} (\theta,\phi) \equiv P_{n}^{prod} A_{np}(\theta) cos(\phi).
\end{equation}

\noindent Measurements and phase shift solutions at intermediate energies \cite{nij} show
the np scattering analyzing power, A$_{np}$, to be a strong function of scattering angle,
but with magnitude no larger than 0.12 over the angle range of interest for the present
experiment. The $\cos (\phi)$ factor reflects the fact that it is only the component of
the vertical neutron polarization perpendicular to the scattering plane for a given np
event that matters.

Since the scattering yield is simply redistributed between scattering toward the left and
the right, there would be no effect at all on cross sections measured with a fully
left-right symmetric forward detector array.  Thus, the only residual polarization effect
changes the measured yield by a fraction:
\begin{equation}
\label{syserr_ii.e_2}
    \delta(\theta) = P_n^{prod} A_{np}(\theta) \int_0^{2\pi} a(\theta,\phi) \cos (\phi) d\phi
    /  \int_0^{2\pi} a(\theta,\phi) d\phi ,
\end{equation}

\noindent where $a (\theta,\phi)$ is the fractional detector acceptance (determined from
fits such as those in Fig.~\ref{fig:acceptance}) in the specified angle bin. The sign
convention used here is that positive $\delta(\theta)$ implies that we observe a higher
event stream 2 yield than we should in the corresponding $\theta_p^{sc}$ bin,
necessitating a correction factor $c_{11} (\theta_p^{sc}) = 1.0 - \delta(\theta)$.

\begin{figure}[htb]
\begin{center}
\includegraphics[scale=0.35]{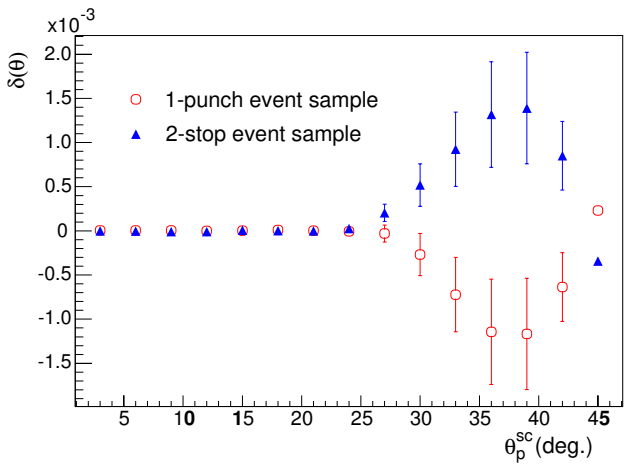}
\caption{\label{fig:poleffect1p}\small The estimated fractional cross section error
introduced by neutron polarization effects for the 1-punch and 2-stop event samples,
plotted as a function of np scattering angle.}
\end{center}
\end{figure}

Figure~\ref{fig:poleffect1p} shows the $\delta(\theta)$ distribution calculated for
1-punch and 2-stop data samples from Eq.~\ref{syserr_ii.e_2}, taking $A_{np}(\theta)$ from
Nijmegen PWA93 calculations \cite{nij}.  We find $\delta = 0$ for both samples at all
angles $\theta_p^{sc} \lesssim 25^\circ$, because the detector has full azimuthal
acceptance in that region.  The small corrections have opposite sign, and hence tend to
cancel, for the two samples at larger angles because the 2-stop events preferentially
populate the left side of the secondary target, while the 1-punch events originate mostly
on the right (see Fig.~\ref{fig:enxtag}). The latter difference is reflected in their
respective $a (\theta,\phi)$ functions (see Fig.~\ref{fig:acceptance}) used in
Eq.~\ref{syserr_ii.e_2}.

While the np analyzing power and the forward detector acceptance functions are well
determined in this experiment, we assign a significant uncertainty to the average tagged
neutron polarization, $P_n^{prod} = -0.10 \pm 0.05$, to account for contributions from
production mechanisms other than charge exchange.  There is correspondingly a $\pm 50\%$
uncertainty assigned to each value of $\delta(\theta)$ in Fig.~\ref{fig:poleffect1p}, but
these errors are completely correlated from one angle bin to another and are strongly
correlated between 2-stop and 1-punch data samples. The largest net uncertainty from
neutron polarization after the (separately corrected) 1-punch and 2-stop results are
combined is $\pm 0.6\times10^{-3}$, and so we conservatively assign an angle-independent
uncertainty of $\pm 0.001$ to $c_{11}$.

\subsection{Target thickness, acceptance and efficiency errors}

\subsubsection{CH$_{2}$ target thickness uncertainty}

The overall normalization uncertainty associated with $t_H$ in Eq.~\ref{diffcrosssec} was
determined to be $\pm 0.4\%$ (\emph{i.e.}, $c_{12} = 1.000 \pm 0.004$) from careful
weighing of the CH$_2$ target used. Because the target only sat in a secondary neutron
beam of low flux, there should not have been any appreciable deterioration in the target
during the length of the run, nor was any visually evident.

\subsubsection{Acceptance uncertainty}

The acceptance uncertainty was determined independently for each angle bin by varying the
most critical one or two detector geometry parameters used in the simulations (see
Fig.~\ref{fig:acceptance}) from their best-fit values until the overall $\chi^2$ value for
the simulated \emph{vs.} measured $\phi_p^{sc}$ distribution in that angle bin increased
by unity. Because the optimized values of $\chi^2$ per degree of freedom for the different
angle bins and event samples were statistically distributed about 1.14, rather than 1.00,
we multiplied these acceptance changes by a uniform factor of 1.07 to arrive at final
systematic uncertainties.  The acceptance uncertainty is strongly angle-dependent, varying
from $\pm 0.001$ at $\theta_{c.m.} > 120^\circ$, where $a_\phi > 95\%$, to $\pm 0.017$ at
$\theta_{c.m.}=90^\circ$, where $a_\phi \approx 50\%$. With this evaluation method, we
consider the estimated uncertainties to be largely uncorrelated from angle bin to bin.

\subsubsection{Wire chamber efficiencies}

The MWPC's were not used at all in forming a hardware trigger, but in software event
reconstruction we required at least one hit registered in each of the x and y planes, for
chambers 1, 2 and 3, plus a $\chi^2$ value below an upper threshold for fitting these hits
to a straight line track. Thus, the overall MWPC efficiency to use in extracting absolute
cross sections is:
\begin{equation}
\label{mwpceff}
    \eta_{MWPC} = \eta_x(1)\eta_y(1) \eta_x(2)\eta_y(2) \eta_x(3)\eta_y(3)
    \eta_{\mbox{fit quality}}.
\end{equation}

\noindent The efficiency of each MWPC plane was determined from tracks reconstructed
without the benefit of the plane in question, based on the fraction of such tracks that
produced a hit on this plane in the immediate vicinity of the reconstructed crossing
point.  Each of the first six factors in Eq.~\ref{mwpceff} was found to exceed 0.99 and
was determined with an uncertainty $\approx \pm 5 \times 10^{-4}$.  Their product is
$0.985 \pm 0.0013$.

The factor $\eta_{\mbox{fit quality}}=0.998 \pm 0.0011$ was determined by estimating the
number of free-scattering events removed from the analysis by the $\chi^2$ cut. This was
done by examining $\Delta E$ spectra for individual angle bins, following carbon
subtraction, for events that failed the fit quality test.  The number of free-scattering
events (and its uncertainty) in each such spectrum was extracted by fitting Gaussian peaks
of the same position and width as those used for the normal $\Delta E$ spectra, such as
Fig.~\ref{fig:bckgrnd_ii.a}(b).  There was no indication in these analyses that any of the
factors in Eq.~\ref{mwpceff} varied with position on the MWPC's, or therefore with
scattering angle.  The overall wire chamber efficiency correction is thus an
angle-independent $c_{14} = 1.0017 \pm 0.002$.

\subsubsection{Trigger inefficiencies}

Inefficiencies in detectors used to form the hardware trigger lead to loss of events in an
unrecoverable way. Possible tagger inefficiencies do not matter here, because they lead to
loss of the same fraction of events from streams 1, 2 and 3, and hence do not affect the
cross sections determined from ratios of event yields in these streams. The two detectors
used to form the hardware trigger for event stream 2, but not for stream 1, are the
$\Delta E$ scintillator and the rear hodoscope. The former was viewed by four phototubes,
at least three of which were required to give signals surpassing threshold in the trigger
logic.  In the data analysis, we were able to determine for each scattering angle bin the
ratio of reconstructed free-scattering events that had only three \emph{vs.} all four
$\Delta E$ phototubes above threshold.  We then estimated the $\Delta E$ trigger
inefficiency under the conservative assumption that the ratio of free-scattering events
with two or fewer phototubes firing to those with three firing would be the same as the
determined ratio of events with three to four firing. (Some illuminated locations on the
scintillator lacked a direct line of sight to one or another, but never simultaneously to
two, of the four phototubes.)  The resulting inefficiency appears to show a systematic
angle-dependence, roughly represented by $0.008 \cos^2 (\theta_p^{sc})$, \emph{i.e.}, the
inefficiency grows as the $\Delta E$ pulse height shrinks.

We have considered two different types of potential hodoscope trigger inefficiencies.
Problems in an individual hodoscope element or phototube would show up as an inefficiency
localized in $\theta$ and $\phi$, and therefore as a deviation of the measured
$\phi$-distribution for some angle bins from the simulated acceptance function. Any such
localized trigger-level inefficiencies should thus be subsumed in the acceptance
uncertainty calculation mentioned above.

However, an electronic inefficiency in the modules forming the hodoscope trigger logic
could have caused equal fractional losses in all angle bins. A limit on this inefficiency
was estimated from event stream 4 (observing tagged protons from the GJT), which included
the Veto2 scintillator directly in front of the hodoscope, but not the hodoscope itself,
in the trigger logic. We found that $(0.6 \pm 0.1)\%$ of these triggered events were not
accompanied by hodoscope signals above threshold in both relevant phototubes, of which
0.4\% have already been accounted for as reaction tail losses below threshold (see
Sec.~\ref{reactiontaillosses}).

Combining the above effects, the overall correction factor for trigger inefficiencies has
been taken as $c_{15} = [1.002+0.008\times cos^2(\theta_p^{sc})] \pm [0.001+0.004\times
cos^2(\theta_p^{sc})]$.  The angle-dependent part of the uncertainty here is intended to
accommodate observed fluctuations in the inferred $\Delta E$ trigger efficiency, and is
viewed as largely uncorrelated among different angle bins.

\subsubsection{Dead time differences among event streams}

Triggers in all event streams were blocked electronically at an early stage in the event
trigger logic by a common busy signal reflecting electronic readout or computer processing
activity in \emph{any} of the event streams.  To first order, then, the different streams
should have a common dead time ($\approx 10\%$ for typical running conditions), and the
dead time should cancel in the event stream ratios from which cross sections are deduced
(see Eq.~\ref{diffcrosssec}).  However, this cancellation is imperfect, as revealed by
ratios of scaler values recording the number of tagged neutron \emph{vs.} tagged proton
candidates before and after busy-vetoing.  Typically, $\approx 1\%$ fewer neutron tags
survived the veto, and this was traced to the occurrence of bursts of electronic noise
triggers from the tagger. While the loss of these noise triggers should not have directly
depleted the valid sample of any event streams (\emph{i.e.}, (2)-(4)) that required other
detectors in coincidence with the tagger, it did reduce the number of valid events
recorded in the neutron flux stream (1), because all raw neutron tags, whether valid or
not, contributed equally to the countdown of a (divide by 20) prescaler used for this
stream. To compensate for this loss of neutron flux events, we have to reduce the
extracted cross sections at all angles by a factor $c_{16} = 0.991 \pm 0.005$.  The
uncertainty in this correction allows for possible model-dependence in our interpretation
of the live-time difference inferred from the scaler ratios.

\subsection{Errors in Determination of Kinematic Variables for np Scattering}

\subsubsection{Neutron energy errors}

As explained in Sec.~\ref{finalresults}, the data were analyzed in narrow neutron energy
slices, with each result then being corrected slightly in order to extract a final overall
cross section at the single mean energy of 194.0 MeV.  There is an overall scale
uncertainty in the tagged neutron energies that we estimate to be $\pm 150$ keV, with
roughly equal contributions from the energy of the stored primary proton beam in the
Cooler and the energies extracted from the tagger for the low-energy recoil protons. The
stored beam energy (202.46 MeV) is based on the precisely measured rf frequency (1.96502
MHz) and the Cooler circumference, which has been previously calibrated \cite{Ste2003} to
better than 1 cm out of 87 m, translating to $\pm 70$ keV. In the ``coasting" (rf off)
mode used for data taking, the beam energy is maintained by interactions with the cooling
electrons, and this may increase the beam energy uncertainty to $\approx 100$ keV. The
energy scale of the recoil protons is calibrated by analysis of $^{228}$Th $\alpha$-source
spectra measured with the tagger \cite{Pet2004}, and its $\pm 100$ keV uncertainty arises
predominantly from thickness uncertainties for detector dead layers, combined with the
quite different corrections for energy loss in the dead layers needed for protons
\emph{vs.} the calibration $\alpha$-particles.

The energy scale uncertainty could be translated into a consequent cross section
uncertainty as a function of angle by using Nijmegen PWA93 calculations to evaluate

\begin{equation}
\delta \sigma_{energy}(\theta_p^{sc}) =  (\pm 150~{\rm keV}) \frac{\partial
[d\sigma/d\Omega] (\theta_p^{sc})}{\partial E}|_{194~{\rm MeV}} .
\end{equation}

Although this systematic error can be angle-dependent, the values at different angles
would still be completely correlated, since the neutron energy scale will be off in the
same direction for all angles. Hence, we prefer to not include this effect in the cross
section uncertainties, but rather to quote the measured cross sections as applying at a
mean neutron energy of $194.0 \pm 0.15$ MeV.

\subsubsection{Scattering angle errors}

A systematic error $\delta \theta_p^{sc}$ in determination of the centroid np scattering
angle within a given analyzed bin is equivalent to an error $\delta \sigma_{angle}
(\theta_p^{sc})$ in the measured differential cross section:

\begin{equation}
\label{syserr_iv.b_1} \delta \sigma_{angle} (\theta_p^{sc}) = \delta \theta_p^{sc}
 \frac{\partial [ d\sigma / d\Omega] (\theta_p^{sc})}{\partial \theta_p^{sc}}|_{194~{\rm MeV}} ,
\end{equation}

\noindent where the angular derivative of the cross section can be taken, for example,
from Nijmegen PWA calculations \cite{nij}. In evaluating $\delta \theta_p^{sc}$, we
consider the contributions from uncertainties $\delta \theta_{inc}$ in the neutron
incidence angle on target deduced from the tagger and $\delta \theta_{out}$ in the angle
of the outgoing proton through the MWPC's: \\

\begin{equation}
\label{syserr_iv.b_2} \delta \theta_p^{sc}= [\langle \delta \theta_{inc} \rangle^2 +
\langle \delta \theta_{out} \rangle^2]^{1/2},
\end{equation}

\noindent where the averages are evaluated over the full free scattering event sample,
over the transverse coordinates $(x_{tag},y_{tag})$ of the scattering origin on the
secondary target, and over all scattering angles.  Consistent values were extracted for
the 2-stop and 1-punch samples.

The angle uncertainties were estimated within $1 \times 1$ cm$^2$ pixels in
$(x_{tag},y_{tag})$ as half the mean event-by-event difference between angles
reconstructed by two different approaches.  In the case of $\theta_{inc}$ one method
utilized tagger information only to predict the neutron trajectory, while the second
considered instead the straight line from the neutron production vertex on the GJT,
inferred from the tagger, to the intersection $(x_{track}, y_{track})$ of the
reconstructed forward proton track with the secondary target.  For $\theta_{out}$ we used
proton tracks reconstructed with MWPC geometry parameters that were either (\emph{i})
optimized to minimize the overall $\chi^2$ value for tracks, or (\emph{ii}) adjusted to
increase overall $\chi^2$ by unity.  A yield-weighted average of the results over all
target pixels gives $\langle \delta \theta_{inc} \rangle = 1.3$ mrad and $\langle \delta
\theta_{out} \rangle = 0.04$ mrad.

The cross sections were not corrected for potential systematic angle errors, but we
extract from Eq.~\ref{syserr_iv.b_1} net systematic uncertainties of $\pm 0.4\%$ for $120
\leq \theta_{c.m.} \leq 180^\circ$, $\pm 0.3\%$ for $100 \leq \theta_{c.m.} \leq
120^\circ$ and $\pm 0.1\%$ for $90 \leq \theta_{c.m.} \leq 100^\circ$.  Because the
extracted incidence angle differences (between the two methods described above) exhibit
sizable fluctuations from one target pixel to another, or from one angle bin to another,
we view these estimated uncertainties as uncorrelated from angle bin to angle bin.

\subsection{Summary of angle-dependence}

The effect of the correction factors $c_i$ associated with the various sources of
systematic error considered in this section is cumulative, and averages 1.10, with small
variations with angle and data sample, as summarized in Table~\ref{tab:syserr}.  We
assume, however, that the various uncertainties are uncorrelated with one another, and we
add them in quadrature to obtain final systematic error estimates.  The majority of error
sources we have considered are explicitly or effectively angle-independent; when combined,
these yield an overall normalization uncertainty of $\pm 1.5\%$.  The uncertainties
associated with our measurements of acceptance and scattering angle, and with trigger
inefficiencies, are considered angle-dependent and uncorrelated from point to point. These
three sources are combined to give the net point-to-point systematic uncertainties in
Table~\ref{tab:finalresults}, where we also collect our final absolute cross section
measurements obtained from a weighted average over the three independently analyzed and
corrected data samples (1-punch, 2-stop with $E_{p1} \leq 5.0$ MeV and 2-stop with $E_{p1}
> 5.0$ MeV). The point-to-point and normalization uncertainties combine to give an overall
systematic error of $\pm 1.6\%$ in most angle bins.

\begin{table}[htp!]
\caption{\label{tab:finalresults} Final differential cross section results for np
scattering at $E_n = 194.0 \pm 0.15$ MeV, averaged over data samples.}
\begin{ruledtabular}
\begin{tabular}{llll}
c.m. angle & $(d\sigma /d \Omega)_{c.m.}$ & Stat. unc. & Syst. unc.\footnote{This column
lists point-to-point systematic uncertainties.  In addition, there is an overall cross
section scale uncertainty of $\pm 1.5\%$.} \\
(deg.)     &    (mb/sr)                   &  (mb/sr)   &   (mb/sr) \\
\hline\hline
92.7    &   1.98   &   0.06  &  0.03 \\
98.8    &   2.00   &   0.05  &  0.02 \\
104.8   &   2.31   &   0.05  &  0.02 \\
110.8   &   2.57   &   0.05  &  0.02 \\
116.8   &   3.01   &   0.05  &  0.02 \\
122.8   &   3.47   &   0.06  &  0.02 \\
128.8   &   4.01   &   0.06  &  0.02 \\
134.9   &   4.75   &   0.07  &  0.03 \\
140.9   &   5.34   &   0.08  &  0.03 \\
146.9   &   5.98   &   0.08  &  0.04 \\
152.9   &   6.64   &   0.10  &  0.04 \\
159.0   &   7.61   &   0.11  &  0.05 \\
165.0   &   8.89   &   0.14  &  0.06 \\
171.0   &  10.62   &   0.19  &  0.07 \\
177.0   &  11.86   &   0.34  &  0.08 \\
\hline
\end{tabular}
\end{ruledtabular}
\end{table}

\section{Discussion}

The preceding section provided a detailed catalogue of the issues that must be carefully
controlled to measure precise absolute cross sections with medium-energy neutron beams. To
our knowledge, no previous experiments have attempted a comparable degree of control. The
best existing absolute neutron-induced cross section standards at intermediate energies
are from attenuation measurements of total cross sections \cite{Lisowski}, which are not
suitable for calibrating neutron fluxes.  It is hoped that the present results will
provide a new calibration standard.  The excellent agreement of our experimentally
determined absolute cross section scale with that given by the Nijmegen PWA93 solution
(see Fig.~\ref{fig:avexs}) confirms the consistency of our results with the total cross
section measurements.

\begin{figure}[htb]
\begin{center}
\includegraphics[scale=0.45,angle=-90]{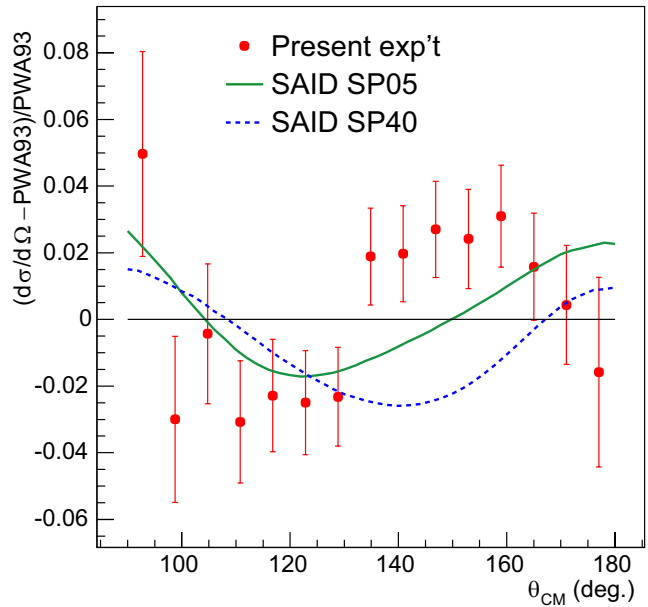}
\caption{\label{fig:xsdif}\small The relative differences of the present absolute np
scattering differential cross sections and of two SAID PWA solutions \cite{SAID,SP40} from
the Nijmegen PWA93 solution \cite{Sto1993,nij}, all at $E_n = 194$ MeV.  The SP40 solution
is from a 2003 analysis of the database from 0--400 MeV, while SP05 is the current SAID
solution, fitted over the range 0--3.0 GeV, including the present data in the fit.}
\end{center}
\end{figure}

The level of agreement of our measurements with PWA's at $E_n = 194$ MeV is presented in
more detail in Fig.~\ref{fig:xsdif}.  Here it is seen that, while the absolute cross
section scale of the experimental results is in very good agreement with the Nijmegen
PWA93 solution, there is a small systematic deviation in angular shape between the two:
our results are higher than PWA93 by 2--3\% for $135 < \theta_{c.m.} < 165^\circ$ and
lower by a similar amount for $100 < \theta_{c.m.} < 130^\circ$.  These deviations
considerably exceed our estimated systematic uncertainty in the angle-dependence.  In
particular, we note that the forward detector acceptance used in the former angle range is
already essentially 100\% (see Fig.~\ref{fig:acceptance2}), so that the extracted cross
section cannot be overestimated by virtue of underestimating acceptance.  Furthermore, the
results for the three independently analyzed data samples agree extremely well in this
angle region (see Fig.~\ref{fig:com1pn2s}).  We do see a possible small, statistically
marginal, systematic deviation among our three data samples in Fig.~\ref{fig:com1pn2s}
over the angle region from 100 to 130$^\circ$, with the 1-punch cross sections falling on
average a few \% below those for the two 2-stop event samples.  However, even if this
difference reflects a real systematic problem, it could only pull the averaged cross
section down by less than 0.5\% in this region, too small to account for the deviation
from PWA93 in Fig.~\ref{fig:xsdif}.

We also show in Fig.~\ref{fig:xsdif} the relative differential cross section differences
between two recent SAID PWA solutions \cite{SAID,SP40} and the Nijmegen PWA93.  The
various PWA solutions differ from one another by as much as 2--3\% also in the angle
region displayed. Furthermore, we note that the SAID solution has shifted by $\sim$2\%
after inclusion of the present results in the fitted np database (even though that
inclusion was carried out by adding our full, mostly angle-independent, systematic
uncertainties in quadrature with our statistical uncertainties, thereby under-weighting
the present data in the fit).  We conclude that the deviations between the present results
and PWA93 are of the same order as the present uncertainties in the PWA solutions, and
most likely point to the need to refit phase shifts.  We note, however, that there is a
conceptual flaw in the procedures for such refitting to a database where all experiments
have systematic uncertainties, but there is considerable variability in the level at which
those systematic uncertainties are reported in the literature.

Finally, we address the comparison of the present results with those from the recent
experiments by the Uppsala \cite{Rah1998} and Freiburg \cite{Fra2000} groups, both of
which have been rejected from the np database used in the Nijmegen and SAID PWA's. As
illustrated in Fig.~\ref{fig:avexs} by the comparison of the two experimental results with
PWA curves at the respective energies of the experiments, the present results deviate
systematically from those of \cite{Rah1998} in the steepness of the back-angle rise in
cross section.  These deviations are larger than the differences anticipated from the
difference in neutron energy between the two experiments.  There is a similar, though not
quite as pronounced, systematic deviation of the present results from those of Franz,
\emph{et al.} \cite{Fra2000}, shown in Fig.~\ref{fig:xscomp}.

\begin{figure}[htb]
\begin{center}
\includegraphics[scale=0.45]{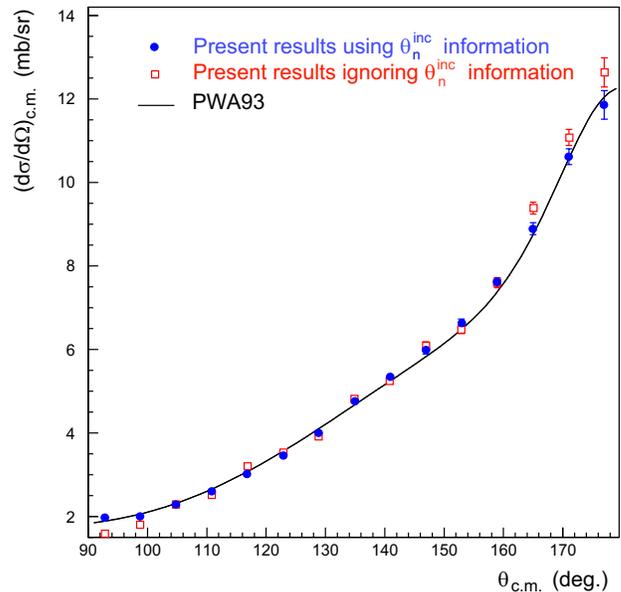}
\caption{\label{fig:xsnoninc}\small The effect on the present analysis of neglecting
tagger information about the neutron incidence angle in the reconstruction of the np
scattering angle for each event.  The closed circles represent the final results, while
the open squares are those when the scattering angle is estimated only with respect to the
central neutron beam direction. Note the suppressed zero on the cross section scale.}
\end{center}
\end{figure}

It is difficult to say definitively whether there might be a common problem that caused
excessive cross sections near $\theta_{c.m.} = 180^\circ$ in both of these earlier,
completely independent and quite different, experiments \cite{{Rah1998},{Fra2000}}.  We
note only that measurements near $\theta_p^{sc} = 0^\circ$, where the solid angle is
vanishing, can be tricky with a secondary neutron beam of sizable angular divergence. One
of the great advantages of the use of a tagged beam is that we are able in the present
experiment to determine the neutron incidence angle event by event. We thereby see that an
angle bin near $\theta_p^{lab} = 0^\circ$ with respect to the \emph{central} neutron
direction in fact includes important contributions from $\theta_p^{sc} > \theta_p^{lab}$
(so that $|d \cos (\theta_p^{sc})|$ for the bin in Eq.~\ref{diffcrosssec} exceeds $|d \cos
(\theta_p^{lab})|$), arising from neutrons deviating from the central beam direction. In
Fig.~\ref{fig:xsnoninc} we compare the present results with those we would have extracted
if we had chosen to ignore the neutron incidence angle information from the tagger in
reconstructing the np scattering angle event by event. Such neglect is seen to give rise
to a systematic overestimate of the cross section at the largest angles by $\sim 5\%$,
comparable to the systematic deviations of the results in Refs.~\cite{{Rah1998},{Fra2000}}
from those of the present experiment. (It also leads to a substantial underestimate near
$\theta_{c.m.} = 90^\circ$, where the yield is falling rapidly with increasing
$\theta_p^{lab}$, due to detector acceptance edges.)  The effect would differ for
different experiments, depending on the angular profile of the neutron beam, including any
effects from scattering off collimator edges (the present experiment used no collimators).

\section{Conclusions}

A tagged intermediate-energy neutron beam produced at the IUCF Cooler ring has facilitated
a measurement of the np scattering differential cross section at 194 MeV bombarding energy
to an absolute precision $\approx \pm 1.5\%$ over the c.m. angular range
90$^\circ$-180$^\circ$. The usage of carefully matched and frequently interchanged solid
CH$_2$ and C secondary targets permitted an accurate background subtraction, reducing
reliance on kinematic cuts that might have introduced larger systematic uncertainties. The
internal consistency in both magnitude and angular shape of the cross sections extracted
from independent data samples characterized by substantially different neutron beam
spatial and energy profiles supports the accuracy of the tagging technique.  Systematic
uncertainties in the measurement, affecting both the overall absolute scale of the cross
sections and the angular dependence, have been carefully delineated, often via auxiliary
measurements and analyses.

The present results are in reasonable agreement with the Nijmegen PWA93 calculation, over
the full angular range covered, although there are systematic deviations at the 2--3\%
level in the angular dependence that might be removed by minor tuning of phase shifts. In
contrast, the present results deviate systematically from other recent measurements
\cite{Rah1998,Fra2000}, especially in the steepness of the back-angle cross section rise.
Our results thus appear to validate the omission of these earlier experiments from the
database used in partial-wave analyses of np elastic scattering, while also suggesting a
possible experimental cause of the earlier overestimates of the cross section near
$\theta_{c.m.} = 180^\circ$.  As the back-angle rise is particularly influential in pole
extrapolations that have occasionally been used \cite{Eri1995,PiN2000} to extract the
charged pion-nucleon-nucleon coupling constant $f_c^2$, the present data also bear on that
coupling strength.  Since our measurements at the largest angles are consistent with, or
even slightly less steep than, the PWA93 solution, a valid pole extrapolation analysis of
the present results should yield a coupling constant value no larger than that ($f_c^2 =
0.0748 \pm 0.0003$) extracted from the Nijmegen PWA \cite{Ren2001}.

\begin{acknowledgments}

We thank the operations staff of the Indiana University Cyclotron Facility for providing
the superior quality cooled beams, and Hal Spinka and Catherine Lechanoine-Leluc for the
loan of critical detector hardware, needed for successful execution of this experiment. We
acknowledge the U.S. National Science Foundation's support under grant numbers
NSF-PHY-9602872, 0100348 and 0457219.

\end{acknowledgments}

\end{document}